\documentclass[11pt,aps,a4paper,eqsecnum,amsmath,amssymb,
     superscriptaddress,notitlepage,nofootinbib]{revtex4-1}

\usepackage{graphicx}

\newcommand{\half}{\frac{1}{2}}

\begin{document}

\title{Finite size properties of staggered $U_q[sl(2|1)]$ superspin chains}

\author{Holger Frahm}

\affiliation{%
Institut f\"ur Theoretische Physik, Leibniz Universit\"at Hannover,
Appelstra\ss{}e 2, 30167 Hannover, Germany}

\author{M\'arcio J. Martins}

\affiliation{%
Departamento de F\'isica, Universidade Federal de S\~ao Carlos,
C.P. 676, 13565-905 S\~ao Carlos (SP), Brazil}

\date{\today}

\begin{abstract}
  Based on the exact solution of the eigenvalue problem for the $U_q[sl(2|1)]$
  vertex model built from alternating 3-dimensional fundamental and dual
  representations by means of the algebraic Bethe ansatz we investigate the
  ground state and low energy excitations of the corresponding mixed superspin
  chain for deformation parameter $q=\exp(-i\gamma/2)$.  The model has a line
  of critical points with central charge $c=0$ and continua of conformal
  dimensions grouped into sectors with $\gamma$-dependent lower edges for
  $0\le\gamma<\pi/2$.  The finite size scaling behaviour is consistent with a
  low energy effective theory consisting of one compact and one non-compact
  bosonic degree of freedom.  In the 'ferromagnetic' regime
  $\pi<\gamma\le2\pi$ the critical theory has $c=-1$ with exponents varying
  continuously with the deformation parameter.  Spin and charge degrees of
  freedom are separated in the finite size spectrum which coincides with that
  of the $U_q[osp(2|2)]$ spin chain.  In the intermediate regime
  $\pi/2<\gamma<\pi$ the finite size scaling of the ground state energy
  depends on the deformation parameter.
\end{abstract}

\maketitle

\section{Introduction}
Studies of exactly solvable two-dimensional vertex models or the equivalent
$(1+1)$-dimensional quantum spin chains can provide important insights into
the nature of excitations in strongly correlated systems and their critical
behaviour.  Over the years this approach has provided much to the present
understanding of such models based on ordinary Lie algebras whose massless
regime is believed to be described by Wess-Zumino-Witten models on the
corresponding group.
On the other hand, many of the physical properties of vertex models based on
Lie \emph{super}algebras and their quantum deformations are still not
understood in detail.
At the same time and in spite of significant progress in recent years only
little is known about the likely candidates for the low energy effective
description of these vertex models, i.e.\ $(1+1)$-dimensional conformal field
theories with non-compact target spaces.  Advances in this direction are
highly desirable as they would likely lead to progress for some problems
related to the duality between gauge and string theories (see
Ref.~\onlinecite{Serb10} and References therein) but also in statistical
mechanics, e.g.\ for the description of disorder driven phase transitions
within the superspin approach to non-interacting electron systems, see e.g.\
\cite{KrOK05,SaSc07}.

One possible approach to this problem is based on the observation that a
non-compact continuum limit can arise from lattice models with a finite number
of states per site \cite{EsFS05,JaSa05,IkJS08}.  If such models are integrable
the powerful techniques of the quantum inverse scattering method allow for a
detailed analysis of their spectrum and ultimately provide important insights
into their continuum limit.  Concerning the possible applications mentioned
above lattice models with alternation between conjugate representations of the
superalgebra have been found to be particular important.  For the integrable
$sl(2|1)$ superspin chain mixing the fundamental representation $3$ and its
dual $\bar{3}$ this approach has led to the identification of the continuum
limit with the $SU(2|1)$ WZW model at level $k=1$ \cite{EsFS05}.
Recently, this has been generalized to find the scattering theory arising in
the continuum limit of the antiferromagnetic $gl(n+N|N)$ spin chains with
$n,N>0$ \cite{Candu10}.

Another direction in which these results may be generalized is by deformation
of the underlying symmetry: in the case of ordinary Lie algebras this has led
to models which exhibit critical lines with anomalous exponents depending
continuously on the deformation parameter.  Whether such a behaviour occurs in
superspin chains when a deformation parameter is introduced and even whether
the critical behaviour of the mixed superspin chain observed in the undeformed
case is robust against the deformation is the question which we want to
address in this paper.

Our paper is organized as follows: below we recall the definition of the mixed
$U_q[sl(2|1)]$ vertex model \cite{Gade99} from which we obtain the integrable
superspin Hamiltonian which is then solved by means of the algebraic Bethe
ansatz.  Since the analysis of the Bethe equations makes use of the known
properties of the Fateev Zamolodchikov model \cite{ZaFa80} with twisted
boundary conditions we also summarize what is known about the different
critical phases of the latter.  In Section~\ref{sec:mix-afm} we present our
results on the low energy properties of the mixed superspin chain in the
'antiferromagnetic' regime which can be interpreted as a regularization of a
continuum theory consisting of a compact and a non-compact boson, similarly to
the $sl(2|1)$ mixed superspin chain discussed in Ref.~\onlinecite{EsFS05}.
The critical behaviour in the 'ferromagnetic' regime discussed in
Section~\ref{sec:mix-fm} turns out to be very different: the corresponding low
energy theory exhibits exact spin-charge separation for all values of the
deformation parameter $q$.  Both the spinon and the holon modes are described
by $U(1)$ Gaussian theories.  In the isotropic limit restoring $sl(2|1)$
invariance the spin part of the spectrum acquires a quadratic dispersion above
an (completely polarized) reference state while the charge part of the
spectrum remains conformal in the same universality class as the isotropic
$osp(2|2)$ spin chain \cite{JaRS03,JaSa05,GaMa07}.

\section{Mixed Vertex Model}

The weights of the mixed vertex model are based on the $R$-matrices associated
to the three dimensional vector representation of $U_q[sl(2|1)]$ and its dual,
labelled $3$ and $\bar{3}$ in the following. These $R$-matrices act on the
tensor products $3\otimes 3$, $3 \otimes \bar{3}$, $\bar{3} \otimes 3$ and $
\bar{3} \otimes \bar{3}$ and we shall denote them by $R^{(33)}(\lambda)$,
$R^{(3 \bar{3})}(\lambda)$, $R^{(\bar{3}, 3)}(\lambda)$ and $R^{(\bar{3},
  \bar{3})}(\lambda)$ respectively. For a general discussion of $R$-matrices
alternating between the vector representation of $U_q[sl(n|m)]$ and its dual
see for instance \cite{ChKu90,DeAk90,BrGZ90,Zhan92}.  In the specific case of
the quantum superalgebra $U_q[sl(2|1)]$ the above set of $R$-matrices have
been explicitly discussed in \cite{Gade99} for a particular grading (models of
this type without grading have been introduced before by Perk and Schultz
\cite{PeSc81}).  In what follows we shall present their expressions for
arbitrary ordering of the Grassmann parities,
\begin{eqnarray}
{\cal R}_{a,b}^{(33)}(\lambda) &=& \sum_{j=1}^{3} {a}_{j}(\lambda) 
e_{j,j}^{(a)} \otimes e_{j,j}^{(b)}
+ b(\lambda) \sum_{\stackrel{j,k=1}{j \neq k}}^{3} 
e_{j,j}^{(a)} \otimes e_{k,k}^{(b)}  
+c(\lambda) \left \{  \sum_{\stackrel{j,k=1}{j > k}}^{3} 
(-1)^{p_{j}p_{k}}e_{j,k}^{(a)} 
\otimes e_{k,j}^{(b)} \right. 
\nonumber \\
&+& \left. \exp(-2\lambda) \sum_{\stackrel{j,k=1}{j < k}}^{3} 
(-1)^{p_{j}p_{k}}  
e_{j,k}^{(a)} \otimes e_{k,j}^{(b)} \right \} ,
\end{eqnarray}

\begin{eqnarray}
{\cal R}_{a,b}^{(3\bar{3})}(\lambda) &=& \sum_{j=1}^{3} {a}_{j}(-\lambda-i\gamma) 
e_{j,j}^{(a)} \otimes e_{j,j}^{(b)}
+ b(-\lambda-i\gamma) \sum_{\stackrel{j,k=1}{j \neq k}}^{3} 
e_{j,j}^{(a)} \otimes e_{k,k}^{(b)} 
\nonumber \\ 
&+&c(-\lambda-i\gamma) \left \{\sum_{\stackrel{j,k=1}{j < k}}^{3} 
(-1)^{p_{j}} q^{-2\delta_{j,1}}e_{j,k}^{(a)} 
\otimes e_{j,k}^{(b)}  
+ \exp(2\lambda+2i\gamma) \sum_{\stackrel{j,k=1}{j > k}}^{3} 
(-1)^{p_{j}} q^{2\delta_{k,1}} 
e_{j,k}^{(a)} \otimes e_{j,k}^{(b)} \right \}, \nonumber \\ 
\end{eqnarray}

\begin{eqnarray}
{\cal R}_{a,b}^{(\bar{3}3)}(\lambda) &=& \sum_{j=1}^{3} {a}_{j}(-\lambda) 
e_{j,j}^{(a)} \otimes e_{j,j}^{(b)}
+ b(-\lambda) \sum_{\stackrel{j,k=1}{j \neq k}}^{3} 
e_{j,j}^{(a)} \otimes e_{k,k}^{(b)}  
+c(-\lambda) \left \{  \sum_{\stackrel{j,k=1}{j > k}}^{3} 
(-1)^{p_{k}} f(k,j)^{-1} e_{j,k}^{(a)} 
\otimes e_{j,k}^{(b)} \right. 
\nonumber \\
&+& \left. \exp(2\lambda) \sum_{\stackrel{j,k=1}{j < k}}^{3} 
(-1)^{p_{k}} f(j,k) 
e_{j,k}^{(a)} \otimes e_{j,k}^{(b)} \right \} ,
\end{eqnarray}

\begin{eqnarray}
{\cal R}_{a,b}^{(\bar{3}\bar{3})}(\lambda) &=& \sum_{j=1}^{3} {a}_{j}(\lambda) 
e_{j,j}^{(a)} \otimes e_{j,j}^{(b)}
+ b(\lambda) \sum_{\stackrel{j,k=1}{j \neq k}}^{3} 
e_{j,j}^{(a)} \otimes e_{k,k}^{(b)}  
+c(\lambda) \left \{  \sum_{\stackrel{j,k=1}{j < k}}^{3} 
(-1)^{p_{j}p_{k}}e_{j,k}^{(a)} 
\otimes e_{k,j}^{(b)} \right. 
\nonumber \\
&+& \left. \exp(-2\lambda) \sum_{\stackrel{j,k=1}{j > k}}^{3} 
(-1)^{p_{j}p_{k}}  
e_{j,k}^{(a)} \otimes e_{k,j}^{(b)} \right \},
\end{eqnarray}
where $e_{j,k}^{(a)} \in \mathrm{End}(\mathbb{C}_{a}^{3})$ are the standard
$3\times 3$ Weyl matrices. The symbol $p_{j}$ denote the Grassmann parities
distinguishing the bosonic $p_{j}=0$ and fermionic $p_{j}=1$ degrees of
freedom.

The dependence of the Boltzmann weights $a_j(\lambda)$, $b(\lambda)$ and
$c(\lambda)$ on the spectral parameter are,
\begin{equation}
a_{j}(\lambda)= \frac{\sinh\left[\lambda-i(2p_j-1)\gamma)\right]}{\sinh\left[\lambda+i\gamma\right]},~~~
b(\lambda)= \frac{\sinh\left[\lambda\right]}{\sinh\left[\lambda+i\gamma\right]},~~~
c(\lambda)= \exp[\lambda]\frac{\sinh\left[i\gamma\right]}{\sinh\left[\lambda+i\gamma\right]},
\end{equation}
while functions $f(j,k)$ depend only on the anisotropy $\gamma$ as follows, 
\begin{equation}
f(1,2)=\exp[2i\gamma (1-p_3)],~~~
f(1,3)=\exp[2i\gamma p_2],~~~
f(2,3)=\exp[-2i\gamma p_1].
\end{equation}

The $R$-matrices defined above fulfill the Yang-Baxter equation on any tensor
product built up from the $3$ and $\bar{3}$ representation, namely
\begin{equation}
R_{12}^{(\omega_1,\omega_2)}(\lambda)
R_{13}^{(\omega_1,\omega_3)}(\lambda+\mu)
R_{23}^{(\omega_2,\omega_3)}(\mu)=
R_{23}^{(\omega_2,\omega_3)}(\mu)
R_{13}^{(\omega_1,\omega_3)}(\lambda+\mu)
R_{12}^{(\omega_1,\omega_2)}(\lambda),
\label{YBE}
\end{equation}
where the representations $\omega_j \in \{ 3, \bar{3} \} $ for $j=1,2,3$.

One consequence of these set of Yang-Baxter relations is that there exists two
different types of Lax operators obeying the Yang-Baxter algebra with the same
$R$-matrix.  For example, this means that an integrable vertex model combining
the $R^{(3,3)}(\lambda)$ and $R^{(3,\bar{3})}(\lambda)$ can be constructed
within the framework of the quantum inverse scattering method.  As usual the
respective row-to-row transfer matrix is written as the supertrace
\cite{Kulish85} over the auxiliary space ${\cal{A}} \sim \mathbb{C}^3$ of the
following ordered product of operators:
\begin{equation}
\label{T33bar}
T^{(3)}(\lambda,\xi) = \mathrm{Str}_{\cal{A}} \left[ 
\mathcal{R}^{(3,3)}_{{\cal A} 2L}(\lambda) 
\mathcal{R}^{(3,\bar{3})}_{{\cal A} 2L-1}(\lambda-i\gamma+\xi) 
\mathcal{R}^{(3,3)}_{{\cal A} 2L-2}(\lambda) \cdots
\mathcal{R}^{(3,\bar{3})}_{{\cal A} 1}(\lambda-i\gamma +\xi) \right]\,
\end{equation}
acting on the Hilbert space $\left(3\otimes \bar{3}\right)^{\otimes L} \sim
\mathbb{C}^{2L}$.
The alternation on the spectral parameter can be introduced since the
$R$-matrices are additive on $\lambda$.  Note that this choice of
inhomogeneity does not spoil the basic properties such as the symmetry and
locality of the interactions of the corresponding alternating superspin chain
with the Hamiltonian obtained by taking the logarithmic derivative of
$T^{(3)}(\lambda,\xi)$ at $\lambda=0$.

By the same token a solvable integrable vertex model alternating
the $R$-matrices $R^{(\bar{3},3)}(\lambda)$ and
$R^{(\bar{3},\bar{3})}(\lambda)$ can be constructed.  The expression of the
transfer matrix mixing such operators is
\begin{equation}
\label{T3bar3}
T^{(\bar{3})}(\lambda,\bar{\xi}) = \mathrm{Str}_{\cal{A}} \left[ 
  \mathcal{R}^{(\bar{3},3)}_{{\cal A} 2L}(\lambda+i\gamma-\bar{\xi}) 
  \mathcal{R}^{(\bar{3},\bar{3})}_{{\cal A} 2L-1}(\lambda) 
  \mathcal{R}^{(\bar{3},3)}_{{\cal A} 2L-2}(\lambda+i\gamma-\bar{\xi}) \cdots
  \mathcal{R}^{(\bar{3},\bar{3})}_{{\cal A} 1}(\lambda) \right].
\end{equation}
acting on the same Hilbert space as (\ref{T33bar}) above.
Again, an alternating superspin chain can be constructed by expansion of the
transfer matrix around $\lambda=0$.

It turns out that -- in addition to commuting among themselves -- the transfer
matrices $T^{(3)}(\lambda,\xi)$ and $T^{(\bar{3})}(\lambda,\bar{\xi})$
constitute a family of commuting operators when the inhomogeneities $\xi$ and
$\bar{\xi}$ are the same, i.e.\
\begin{equation}
\label{comut}
\left [
T^{(3)}(\lambda,\xi), 
T^{(\bar{3})}(\mu,\xi) \right ]=0, \quad\forall\, \lambda,\mu\, .
\end{equation}
The property (\ref{comut}) relies on the fact that we have chosen an identical
ordering of representations $3$ and $\bar{3}$ in the definition of the Hilbert
spaces for the transfer matrices (\ref{T33bar}) and (\ref{T3bar3}) and follows
from the Yang-Baxter equation (\ref{YBE}) once we choose the representations
$\omega_1=3$, $\omega_2=\bar{3}$ and $\omega_3=3,\bar{3}$.  In this situation
we are able to construct an integrable vertex model that alternates
representations $3$ and $\bar{3}$ both on horizontal \emph{and} vertical
spaces of states of a square lattice of size $2L \times 2L$.  The respective
'double row' transfer matrix of such model is obtained by taking the following
product,
\begin{equation}
\label{Tmixed}
  T^{(mix)}(\lambda,\xi)=T^{(3)}(\lambda,\xi) T^{(\bar{3})}(\lambda,\xi)\, .
\end{equation}
By construction $T^{(mix)}(\lambda=0,\xi)$ is proportional to the translation
operator by two lattice sites.  Therefore, we can define an integrable
superspin  Hamiltonian,
\begin{equation}
\label{hammix}
  \mathcal{H}^{(mix)} = i\,\left. \frac{\partial}{\partial\lambda}
    \ln\,T^{(mix)}(\lambda,\xi) \right|_{\lambda=0}\,. 
\end{equation}
The expression for $H^{(mix)}$ in terms of the $R$-matrices can be obtained by
computing the individual Hamiltonians associated with the transfer matrices
$T^{(3)}(\lambda)$ and $T^{(\bar{3})}(\lambda)$.  
The technicalities entering this computation are cumbersome but the final
result is somehow simple,
\begin{eqnarray}
{\cal H}^{(mix)}&=i&\sum_{\stackrel{j=2}{\text{mod~2}}}^{2L} 
\left[{\cal R}^{(3,\bar{3})}_{j,j+1}(\xi-i\gamma) \right]^{-1} 
\left [{\cal \dot{R}}^{(3,\bar{3})}_{j,j+1}(\xi-i\gamma) + 
P_{j,j+2} {\cal \dot{R}}^{(3,3)}_{j,j+2}(0) {\cal R}^{(3,\bar{3})}_{j,j+1}(\xi-i\gamma)
\right ]
\nonumber \\
&+&i\sum_{\stackrel{j=1}{\text{mod~2}}}^{2L} 
\left[{\cal R}^{(\bar{3},3)}_{j,j+1}(i\gamma-\xi) \right]^{-1} 
\left [{\cal \dot{R}}^{(\bar{3},3)}_{j,j+1}(i\gamma-\xi) + 
P_{j,j+2} {\cal \dot{R}}^{(\bar{3},\bar{3})}_{j,j+2}(0) {\cal R}^{(\bar{3},3)}_{j,j+1}(i\gamma-\xi)
\right ]
\end{eqnarray}
where periodic boundary conditions $2L+1 \equiv 1$ and $2L+2 \equiv 2$ are assumed. The
operator $P_{ab}$ is the graded permutation $\displaystyle
P_{ab}=\sum_{j,k=1}^{3} (-1)^{p_{j}p_{k}} {e}_{j,k}^{(a)} \otimes
{e}_{k,j}^{(b)}$ and ${\cal \dot{R}}_{ab}(\lambda)$ denotes the derivative of
the $R$-matrix ${\cal R}_{ab}(\lambda)$ with respect to the spectral parameter
$\lambda$.

The diagonalization of the above transfer matrix can be carried out by
applying the nested algebraic Bethe ansatz approach \cite{KuRe83,BaVV82}.  For
this particular mixed vertex model the essential tools to obtain the
eigenvalues of $T^{(mix)}(\lambda,\xi)$ can for instance be found in
\cite{RiMa06}.  We shall not repeat here these technical details and
concentrate our attention only to the main results.  As usual the expressions
for the eigenvalues obtained in this approach will depend on the choice of
grading \cite{Kulish85,EsKo92,FoKa93,PfFr96,PfFr97}.  For later convenience we
use $[p_1,p_2,p_3]=[0,1,0]$ in the following.
Let $\Lambda_{N_1,N_2}^{(mix)}(\lambda)$ denote the eigenvalues of
$T^{(mix)}(\lambda,\xi)$ in the sector of the Hilbert space selected by fixing
the two conserved quantum numbers related to the $U(1)$ subalgebras of
$U_q[sl(2|1)$, i.e.\ charge $b=(N_1-N_2)/2$ and $z$-component of the spin
$s_3=L-(N_1+N_2)/2$.  As a consequence of (\ref{comut}) the eigenvalues can be
factorized
\begin{equation}
\Lambda_{N_1,N_2}^{(mix)}(\lambda)=
\Lambda^{(3)}_{N_1,N_2}(\lambda)
\Lambda^{(\bar{3})}_{N_1,N_2}(\lambda)\,.
\end{equation}
Here $\Lambda^{(3)}_{N_1,N_2}(\lambda)$ and
$\Lambda^{(\bar{3})}_{N_1,N_2}(\lambda)$ are the corresponding eigenvalues
associated to the transfer matrices $T^{(3)}(\lambda,\xi) $ and
$T^{(\bar{3})}(\lambda,\xi) $ respectively.

It turns out that the expressions for the eigenvalues
$\Lambda^{(3)}_{N_1,N_2}(\lambda)$ and
$\Lambda^{(\bar{3})}_{N_1,N_2})(\lambda)$ are given by,
\begin{eqnarray}
\label{GAMAMIX1}
\Lambda^{(3)}_{N_1,N_2}(\lambda)&=&
\left[\frac{\sinh(\lambda+\xi)}{\sinh(\lambda+\xi-i\gamma)}\right]^{L}  
\prod_{j=1}^{N_{1}} \frac{\sinh( \lambda_{j}^{(1)}-\lambda +i\gamma/2)}
{\sinh( \lambda_{j}^{(1)}-\lambda -i\gamma/2)} 
+\left[\frac{\sinh(\lambda)}{\sinh(\lambda+i\gamma)}\right]^{L}  
\prod_{j=1}^{N_{2}} \frac{\sinh(\lambda- \lambda_{j}^{(2)}+ i\gamma)}
{\sinh(\lambda- \lambda_{j}^{(2)})} \nonumber \\
&-&
\left[\frac{\sinh(\lambda+\xi)\sinh(\lambda)}{\sinh(\lambda+i\gamma)\sinh(\lambda+\xi-i\gamma)}\right]^{L}  
\prod_{j=1}^{N_{1}} \frac{\sinh(\lambda- \lambda_{j}^{(1)} -i\gamma/2)}
{\sinh( \lambda- \lambda_{j}^{(1)}+i\gamma/2)} 
\prod_{j=1}^{N_{2}} \frac{\sinh( \lambda_{j}^{(2)}-\lambda -i\gamma)}
{\sinh( \lambda_{j}^{(2)}-\lambda) }, \nonumber \\
\end{eqnarray}
and
\begin{eqnarray}
\label{GAMAMIX2}
\Lambda^{(\bar{3})}_{N_1,N_2}(\lambda)&=&
\left[\frac{\sinh(\lambda)}{\sinh(\lambda+i\gamma)}\right]^{L}  
\prod_{j=1}^{N_{1}} \frac{\sinh(\lambda- \lambda_{j}^{(1)}-\xi+3i\gamma/2)}
{\sinh( \lambda- \lambda_{j}^{(1)} -\xi +i\gamma/2)} 
\nonumber \\
&+&\left[\frac{\sinh(\lambda-\xi+i\gamma)}{\sinh(\lambda-\xi)}\right]^{L}  
\prod_{j=1}^{N_{2}} \frac{\sinh(\lambda_{j}^{(2)}-\lambda+ \xi)}
{\sinh(\lambda_{j}^{(2)}-\lambda+\xi-i\gamma)} \nonumber \\
&-&
\left[\frac{\sinh(\lambda-\xi+i\gamma)\sinh(\lambda)}{\sinh(\lambda+i\gamma)\sinh(\lambda-\xi)}\right]^{L}  
\prod_{j=1}^{N_{1}} \frac{\sinh(\lambda_{j}^{(1)}-\lambda +\xi -3i\gamma/2)}
{\sinh(\lambda_{j}^{(1)}-\lambda+\xi-i\gamma/2)} 
\prod_{j=1}^{N_{2}} \frac{\sinh( \lambda- \lambda_{j}^{(2)}-\xi)}
{\sinh(\lambda- \lambda_{j}^{(2)}-\xi+i\gamma) }, \nonumber \\
\end{eqnarray}
where the rapidities $\lambda_{j}^{(1)}$ and $\lambda_{j}^{(2)}$ satisfy the
following set of nested Bethe equations,
\begin{equation}
\label{betheM}
\begin{aligned}
 \left[\frac{\sinh(\lambda_{j}^{(1)}+i\gamma/2)}
 {\sinh(\lambda_{j}^{(1)}-i\gamma/2)}\right]^{L}&=
 \prod_{k=1}^{N_{2}}
  \frac{\sinh(\lambda_{j}^{(1)}-\lambda_{k}^{(2)}+i\gamma/2)}
 {\sinh(\lambda_{j}^{(1)}-\lambda_{k}^{(2)}-i\gamma/2)},\quad j=1,\cdots,N_1 , 
\\
 \left[\frac{\sinh(\lambda_{j}^{(2)}+\xi)}
 {\sinh(\lambda_{j}^{(2)}+\xi-i\gamma)}\right]^{L}&=
 \prod_{k=1}^{N_{1}}
   \frac{\sinh(\lambda_{j}^{(2)}-\lambda_{k}^{(1)}+i\gamma/2)}
 {\sinh(\lambda_{j}^{(2)}-\lambda_{k}^{(1)}-i\gamma/2)},\quad j=1,\cdots,N_2 .
\end{aligned}
\end{equation}
Note that for the particular choice $\xi =i\gamma/2$ these Bethe ansatz
equations become symmetrical on the variables $\lambda_j^{(1)}$ and
$\lambda_j^{(2)}$.  In the following we will concentrate our studies on this
case.  The eigenvalues of the Hamiltonian (\ref{hammix}) corresponding to a
solution of (\ref{betheM}) are
\begin{equation}
\label{ener-sl}
\begin{aligned}
  E^{(mix)}_{N_1,N_2}(\gamma) &= i\,\left. \frac{\partial}{\partial\lambda}
   \ln\,\Lambda_{N_1,N_2}^{(mix)}(\lambda,i\gamma/2) \right|_{\lambda=0}\, 
\\
   &= 4L\cot\frac{\gamma}{2}
   + 2\left(\sum_{k=1}^{N_1}
     \frac{\sin\gamma}{\cos\gamma-\cosh2\lambda_k^{(1)}}
   + \sum_{k=1}^{N_2}
   \frac{\sin\gamma}{\cos\gamma-\cosh2\lambda_k^{(2)}}\right)\,.  
\end{aligned}
\end{equation}
Considering the above solution one sees that the spectrum at the points
$\gamma$ and $2 \pi -\gamma$ are related to each other by only a sign,
\begin{equation}
\label{relation}
E^{(mix)}_{N_1,N_2}(\gamma) = -E^{(mix)}_{N_1,N_2}(2\pi- \gamma) \,.
\end{equation}

\section{The twisted XXZ spin-1 model }
The classical vertex model associated to the integrable Heisenberg XXZ spin-1
chain turns out to be the three-state factorized $R$-matrix found by
Zamolodchikov and Fateev \cite{ZaFa80}. Considering our previous notation this
operator can be expressed as
\begin{equation}
\label{XXZmatrix}
\begin{aligned}
\mathcal{R}_{ab}(\lambda) & = e_{1,1}^{(a)} \otimes e_{1,1}^{(a)}
+ e_{3,3}^{(a)} \otimes e_{3,3}^{(b)}
+\bar{f}(\lambda) \left [ e_{1,1}^{(a)} \otimes e_{3,3}^{(b)} +
e_{3,3}^{(a)} \otimes e_{1,1}^{(b)} \right ],\\
&+\bar{b}(\lambda) \left [ e_{1,1}^{(a)} \otimes e_{2,2}^{(b)} +
e_{2,2}^{(a)} \otimes e_{1,1}^{(b)}+ 
e_{2,2}^{(a)} \otimes e_{3,3}^{(b)} 
+ e_{3,3}^{(a)} \otimes e_{2,2}^{(b)} 
+ e_{2,2}^{(a)} \otimes e_{2,2}^{(b)} 
\right ]\\
&+\bar{c}(\lambda) \left[ e_{1,2}^{(a)} \otimes e_{2,1}^{(b)} +
e_{2,1}^{(a)} \otimes e_{1,2}^{(b)} 
+e_{2,3}^{(a)} \otimes e_{3,2}^{(b)} +
e_{3,2}^{(a)} \otimes e_{2,3}^{(b)} \right ] \\
&+\bar{d}(\lambda)\left [ e_{1,2}^{(a)} \otimes e_{3,2}^{(b)} 
+ e_{2,3}^{(a)} \otimes e_{2,1}^{(b)} 
+e_{2,1}^{(a)} \otimes e_{2,3}^{(b)} 
+ e_{3,2}^{(a)} \otimes e_{1,2}^{(b)} \right ]\\
&+ \bar{h}(\lambda)\left[ e_{1,3}^{(a)} \otimes e_{3,1}^{(b)}
+e_{2,2}^{(a)} \otimes e_{2,2}^{(b)}
+e_{3,1}^{(a)} \otimes e_{1,3}^{(b)} \right]
\end{aligned}
\end{equation}
where the corresponding Boltzmann weights $\bar{b}(\lambda)$,
$\bar{c}(\lambda)$,
$\bar{d}(\lambda)$,
$\bar{f}(\lambda)$
and $\bar{h}(\lambda)$ are given by,
\begin{equation}
\label{weightXXZbc}
\begin{aligned}
&\bar{b}(\lambda) = \frac{\sinh(\lambda)}{\sinh{(\lambda+i\gamma)}},\quad
\bar{c}(\lambda) = \frac{\sinh(i\gamma)}{\sinh(\lambda+i\gamma)},\quad
\bar{d}(\lambda) = \frac{\sinh(i\gamma)\sinh(\lambda)}{\sinh(\lambda+i\frac{\gamma}{2})\sinh(\lambda+i\gamma)},
\\
&\bar{f}(\lambda) =
\frac{\sinh(\lambda-i\frac{\gamma}{2})\sinh(\lambda)}{\sinh(\lambda+i\frac{\gamma}{2})\sinh(\lambda+i\gamma)},\quad 
\bar{h}(\lambda) =
\frac{2\cosh(\frac{i\gamma}{2})\left[\sinh(i\frac{\gamma}{2})\right]^2} 
{\sinh(\lambda+i\frac{\gamma}{2})\sinh(\lambda+i\gamma)} .
\end{aligned}
\end{equation}

With (\ref{XXZmatrix}) the respective transfer-matrix $T(\lambda)$ with
toroidal boundary conditions can formally be constructed as follows
\begin{equation}
\label{TXXZ}
T(\lambda) = Tr_{\cal{A}} \left[ G_{\cal A}
\mathcal{R}_{{\cal A} L}(\lambda) \mathcal{R}_{{\cal A} L-1}(\lambda) \cdots
\mathcal{R}_{{\cal A} 1}(\lambda) \right],
\end{equation}
where $G_{\cal A}$ denotes a $ 3 \times 3 $ matrix representing the twisted
boundary condition. 

The diagonal twisted boundary condition compatible with integrability is
obtained by choosing the matrix $G$ as,
\begin{equation}
G_{\cal A}=\left(\begin{array}{ccc}
        \,1\, & 0 & 0 \\
        0 & \mathrm{e}^{i\varphi} &  0 \\
        0 & 0 & \mathrm{e}^{2i\varphi} \\
        \end{array}\right),
\label{bound}
\end{equation}
where the angle $\varphi$ is assumed to be in the interval $ 0 \leq \varphi \leq
\pi$. 

The transfer matrix (\ref{TXXZ}) with (\ref{bound}) can be diagonalized with
very little difference from the periodic case since the presence of the
diagonal boundary matrix $G_{\cal{A}}$ preserves the $U(1)$ bulk symmetry.
The respective eigenvalues can be determined either by using the mechanism of
fusion \cite{BaTs86,KiRe86,KiRe87} or by applying the algebraic Bethe ansatz
construction developed in \cite{MeMa09}.  As a consequence of the $U(1)$
invariance of the transfer matrix the Hilbert space can be separated in
disjoint sectors corresponding to total magnetization $s_3$.  Starting from
the state with maximal $s_3=L$ one obtains the following expression of the
corresponding eigenvalues $\Lambda_{N}(\lambda,\varphi)$ in the sector
$s_3=L-N$, $N=0,\cdots,L$ 
\begin{equation}
\label{GAMAXXZ}
\begin{aligned}
\Lambda_{N}(\lambda,\varphi) &=
\prod_{j=1}^{N} \frac{\sinh(\lambda_{j}-\lambda+i\gamma/2)}
{\sinh(\lambda_{j} -\lambda -i\gamma/2)} 
+\mathrm{e}^{2i\varphi}
    \left[\frac{\sinh(\lambda-i\gamma/2)\sinh(\lambda)}{ 
                \sinh(\lambda+i\gamma)\sinh(\lambda+i\gamma/2)}\right]^{L}  
    \prod_{j=1}^{N} \frac{\sinh(\lambda-\lambda_{j}+i\gamma)}{
                        \sinh(\lambda-\lambda_j)} \\ 
&+\mathrm{e}^{i\varphi}
    \left[\frac{\sinh(\lambda)}{\sinh(\lambda+i\gamma)}\right]^{L}  
    \prod_{j=1}^{N} \frac{\sinh(\lambda- \lambda_{j} +i\gamma)
                        \sinh(\lambda-\lambda_j-i\gamma/2)}{
          \sinh(\lambda- \lambda_{j}+i\gamma/2) \sinh(\lambda-\lambda_j)},
\end{aligned}
\end{equation}
where the rapidities $\lambda_j$ satisfy the following Bethe ansatz equations,
\begin{equation}
\label{betheXXZ}
\left[\frac{\sinh(\lambda_{j}+i\gamma/2)}
{\sinh(\lambda_{j}-i\gamma/2)}\right]^{L}=
\mathrm{e}^{i\varphi}
\prod_{\stackrel{k=1}{k \neq j}}^{N}  \frac{\sinh(\lambda_{j}-\lambda_{k}+i\gamma/2)}
{\sinh(\lambda_{j}-\lambda_{k}-i\gamma/2)},\quad j=1,\cdots,N .
\end{equation}

At this point we have gathered the basic ingredients allowing to establish a
mapping among part of the spectrum of the mixed transfer matrix (\ref{Tmixed})
and the eigenvalues of the XXZ spin-$1$ model with special toroidal boundary
condition for the special choice of $\xi =i\gamma/2$ in the transfer matrix
$T^{(mix)}(\lambda)$ of the mixed spin chain:
as mentioned above the Bethe ansatz equations (\ref{betheM}) become symmetric
between the two levels for this choice of $\xi$ and their structure resembles
that of the Bethe ansatz equations (\ref{betheXXZ}) of the $XXZ-1$ chain with
twist $\varphi=\pi$.
In particular we find that in the sector $N_1=N_2=N$ of the Hilbert space of
the mixed chain (this is where the total $U_q[sl(2|1)]$ charge $b$ of the
Bethe state is zero) there exists a subset of eigenstates parametrized by
Bethe roots which can be identified with eigenstates of the spin-$1$ chain in
the sector $s_3=L-N$ by setting $\lambda_j^{(1)}=\lambda_j^{(2)} \equiv
\lambda_j$.  A similar correspondence has been observed between a subset
of eigenvalues of the alternating $sl(2|1)$ superspin chain obtained in the
limit $\gamma\to0$ from the mixed chain considered here and the
$SU(2)$-invariant spin-$1$ Takhtajan-Babujian chain \cite{EsFS05}.  For such
states, a direct inspection of the expressions for the eigenvalues
(\ref{GAMAMIX1}), (\ref{GAMAMIX2}), (\ref{GAMAXXZ}) leads us to the following
relation,
\begin{equation}
\label{transf-map}
\Lambda_{N,N}^{(mix)}(\lambda)= 
\left[\frac{\sinh(\lambda+i\gamma/2)}
{\sinh(\lambda-i\gamma/2)}\right]^{2L}
\left[ \Lambda_{N}(\lambda,\varphi=\pi) \right ]^2 .
\end{equation}
As a consequence, we obtain from (\ref{ener-sl}) a relation between the energy
eigenvalues of the XXZ spin-1 chain with $L$ sites and those of the
$U_q[sl(2|1)]$ superspin chain with $2L$ sites of alternating representations
$3$ and $\bar3$:
\begin{equation}
\label{ener-map}
 E_{N,N}^{(mix)} = 4L\cot(\gamma/2) + 2 E_N^{(XXZ)}(\varphi=\pi)
\end{equation}
where
\begin{equation}
\label{ener-ft}
 E_N^{(XXZ)}(\varphi) = i\,\left. \frac{\partial}{\partial\lambda}
   \ln\,\Lambda_{n}(\lambda,\varphi) \right|_{\lambda=0}\, 
 = \sum_{k=1}^{N}
     \frac{\sin\gamma}{\cos\gamma-\cosh2\lambda_k}\,
\end{equation}
is the energy eigenvalue of the XXZ spin-$1$ chain corresponding to a solution
of the Bethe equations (\ref{betheXXZ}).  As in Eq.~(\ref{relation}) for the
mixed superspin chain the spectrum of the XXZ model is inverted at anisotropy
$\gamma=\pi$, i.e.\ $\mathrm{spec}(\gamma) \leftrightarrow
-\mathrm{spec}(2\pi-\gamma)$.

%
%

In the thermodynamic limit $L\to\infty$ the solutions of the Bethe equations
(\ref{betheXXZ}) are grouped into 'strings' consisting of $m$ complex
rapidities $\lambda_j^{(m)}$ characterized by a common real center
$\lambda^{(m)}$ and a parity $v_m=\pm1$:
\begin{equation}
\label{stringXXZ}
  \lambda_j^{(m)} = \lambda^{(m)} +
  i\frac{\gamma}{4}\left({m+1}-2j\right)
  + i\frac{\pi}{4}\,(1-v_m)\,,\quad j=1,\ldots,m\,.
\end{equation}
The allowed values of $(m,v_m)$ depend on the anisotropy $\gamma$ in an
involved way \cite{TaSu72,KiRe86}.  Here we shall not go into the details of
the string classification: for the range of anisotropies $0\le\gamma<\pi$ that
we are considering in this paper it is found that the most relevant root
configurations solving (\ref{betheXXZ}) can be organized into strings $(1,+)$,
$(1,-)$ and $(2,+)$.

\subsection{The disordered antiferromagnet regime of the spin-1 chain}
\label{sec:xxz-afm}
Most importantly, the ground state of the system without twist, $\varphi=0$, and
on an even length lattice is a condensate of $L/2$ $(2,+)$-strings in this
regime \cite{Sogo84,BaTs86,KiRe86}.  In the thermodynamic limit one can
compute the energy per site giving $\epsilon_\infty(\gamma) =
-2\cot({\gamma}/{2})$. The finite size spectrum of the XXZ spin-$1$ model
without twist has been investigated in \cite{AlMa89,FrYF90}: in the entire
interval $0\le\gamma<\pi$ the spectrum has gapless excitations with Fermi
velocity $v_F = {2\pi}/{\gamma}$.  The central charge of the conformal field
theory describing the low energy sector is $c=3/2$, hence the ground state
energy for even $L$ scales as
\begin{equation}
\label{e0-cft}
  E^{(XXZ)}(\varphi=0) - L\epsilon_\infty(\gamma)
  = -\frac{\pi v_F}{6L}\,c + o(\frac{1}{L})
  = -\frac{\pi v_F}{4L} + o(\frac{1}{L})\,.
\end{equation}
The operators of this CFT are given by products of Ising operators and $U(1)$
Kac-Moody fields.  The scaling dimensions of these composite fields in the
presence of a twist $\varphi$ are
\cite{AlMa90}
\begin{equation}
\label{fse-cft}
\begin{aligned}
  &X_{(r,j)}^{(n,m+\varphi/\pi)} (\gamma) =  X_I(r,j) +
    n^2 X_c + \left(m+\frac{\varphi}{\pi}\right)^2 \,
    \frac{1}{16\,X_c}\,,\quad
  X_c= \frac{\pi-\gamma}{4\pi}\,,\\
 &X_I(0,0) \in\{0,1\}\,,\quad
  X_I(0,1) = X_I(1,0) = \frac{1}{8}\,,\quad
  X_I(1,1)=\frac{1}{2}\,.
\end{aligned}
\end{equation}
Depending on the parity of $L$ the possible subset of the KM representations
is determined by the selection rules
\begin{equation}
\begin{aligned}
 n=r+L\,\,\mathrm{mod}\,2\,,\quad
 m=j+L\,\,\mathrm{mod}\,2\,
\end{aligned}
\end{equation}
for given parity $j$  and toroidal b.c. (type $r$) of the Ising sector.
The smallest exponents obtained from (\ref{fse-cft}) for even $L$ are
\begin{equation}
\label{xxzcfteven}
\begin{aligned}
X_{(0,0)}^{(0,0+\varphi/\pi)} (\gamma) &= 
     \left(\frac{\varphi}{\pi}\right)^2 \,
    \frac{1}{16\,X_c}\,,\\
X_{(0,1)}^{(0,-1+\varphi/\pi)} (\gamma) &= \frac{1}{8} +
    \left(1-\frac{\varphi}{\pi}\right)^2 \,
    \frac{1}{16\,X_c}\,,\\
X_{(1,0)}^{(1,0+\varphi/\pi)} (\gamma) &= \frac{1}{8} +
    X_c + \left(\frac{\varphi}{\pi}\right)^2 \,
    \frac{1}{16\,X_c}\,,
\end{aligned}
\end{equation}
and for $L$ odd
\begin{equation}
\label{xxzcftodd}
\begin{aligned}
X_{(1,0)}^{(0,-1+\varphi/\pi)} (\gamma) &= \frac{1}{8} +
    \left(1-\frac{\varphi}{\pi}\right)^2 \,
    \frac{1}{16\,X_c}\,,\\
X_{(0,0)}^{(1,-1+\varphi/\pi)} (\gamma) &= 
    X_c + \left(1-\frac{\varphi}{\pi}\right)^2 \,
    \frac{1}{16\,X_c}\,,\\
X_{(0,1)}^{(1,0+\varphi/\pi)} (\gamma) &= \frac{1}{8} +
    X_c + \left(\frac{\varphi}{\pi}\right)^2 \,
    \frac{1}{16\,X_c}\,.
\end{aligned}
\end{equation}

\begin{figure}[t]
\includegraphics[width=0.75\textwidth]{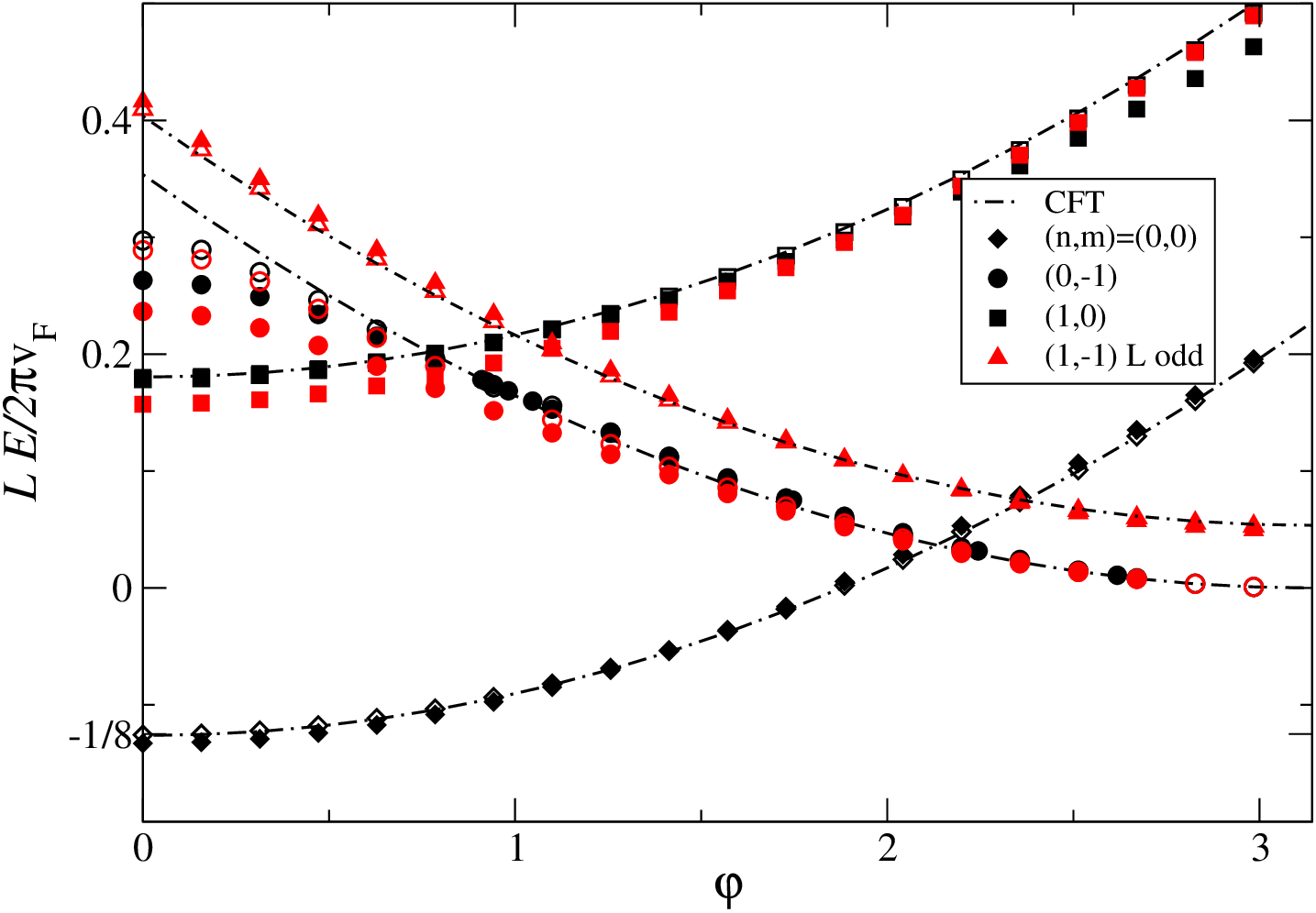}
\caption{\label{fig:fz-tw} Evolution of the ground state and lowest excitation
  energies of the XXZ spin-1 model with $\gamma=\frac{2\pi}{7}$ as a function
  of twist $\varphi$.  Filled (open) symbols in black are for $L=6$
  ($30$), respectively; red symbols for $L=7$ ($31)$.  Dash-dotted lines
  indicate the CFT predictions (\ref{fse-cft}). }
\end{figure}

As the twist is varied the energy of the $\varphi=0$ ground state ($(r,j)=(0,0)$,
$(m,n)=(0,0)$ for even $L$) increases until there occurs a crossing with the
level evolving from $(n,m)=(0,-1)$  at $\varphi=(3\pi-\gamma)/4$ (see
Fig.~\ref{fig:fz-tw}).
The conformal dimension of this state at twist $\varphi=\pi$ is
$X_{0,1}^{0,0}=1/8$ for even and $X_{1,0}^{0,0}=1/8$ for odd length lattices
independent of the deformation parameter $\gamma$.  Together with the finite
size scaling of the ground state energy (\ref{e0-cft}) this gives a state
whose energy is $L\epsilon_\infty(\gamma)$ \emph{without any finite size
  corrections}!  Note that the corresponding wave function does vary with
$\gamma$.
According to (\ref{ener-ft}) this can only be realized with a highly
degenerate configuration of Bethe roots, namely $\lambda_k\equiv 0$ for all
$k=1,\ldots, L$.  The identification (\ref{ener-map}) implies the existence of
a zero energy eigenstate of the mixed $U_q[sl(2|1)]$ superspin model in the
singlet sector of the latter.
For $0\le\gamma\le\pi/2$ this is the ground state of the $\pi$-twisted XXZ
spin-1 chain and the mixed superspin chain, implying that the effective
central charges of this models are $c=0$.  For $\gamma>\pi/2$, the two-fold
degenerate level with scaling dimension $X_{0,0}^{1,-1+\varphi/\pi}$ realized in
the XXZ spin-1 chain of odd length has a lower energy at twist $\varphi=\pi$ .  The
consequences of this crossing for the superspin chain will be discussed below.
%
\subsection{The disordered ferromagnetic regime of the spin-1 chain}
\label{sec:xxz-fm}
As a consequence of the inversion of the spectrum under $\gamma
\leftrightarrow \tilde{\gamma} \equiv 2\pi-\gamma$ we can discuss the
properties of the spin-1 chain in this regime in the same interval
$0<\gamma<\pi$ while changing the sign of the energies (\ref{ener-ft}).  This
leaves the string classification (\ref{stringXXZ}) unchanged.  Without twist
in the boundary conditions the configuration of Bethe roots corresponding to
the ground state in the disordered ferromagnetic regime is given by a filled
sea of $L$ $(1,-)$-strings.
Above this state there are gapless low energy excitations with Fermi velocity
$\tilde{v}_F=2\pi/(2\pi-\gamma)$.  The corresponding conformal field theory
has been identified as a $U(1)$ Gaussian model with central charge $c=1$ and
scaling dimensions of the primary operators given by \cite{AlMa89a}
\begin{equation}
\label{fse-fm}
  \tilde{X}_{n,m}(\gamma) = n^2 x_p + \frac{m^2}{4x_p}\,,
     \qquad x_p = \frac{\gamma}{4\pi}\,
\end{equation}
where $n$ and $m$ take integer values which determine the magnetization
$s_3=n$ and vorticity of the corresponding state.  Note that the
compactification radius of the boson $R=\sqrt{x_p}$ vanishes as $\gamma\to0$
indicating the transition into the non-conformal isotropic ferromagnetic
state.
For toroidal boundary conditions with twist $\varphi$ one has to replace $m\to
m+\varphi/2\pi$ in Eq.~(\ref{fse-fm}).  The adiabatic evolution of the Bethe
roots under the twist is rather involved, a detailed study of the
corresponding regime in the spin-1/2 XXZ chain can be found in
Ref.~\onlinecite{YuFo92}.  

Here our focus is on the antiperiodic twisted chain: choosing $\varphi=\pi$
implies that the finite size gaps are given by Eq.~(\ref{fse-fm}) with integer
$n$ but half-odd integer $m=\pm1/2,\pm3/2,\ldots$.  As an immediate
consequence the lowest state of the conformal part of the spectrum in the
finite system has an energy ($\tilde{\epsilon}_\infty^{(XXZ)}$ is the bulk ground
state energy density of the chain in this regime)
\begin{equation}
  E_0(L)-L\tilde{\epsilon}_\infty^{(XXZ)} \simeq 
      -\frac{\pi\tilde{v}_F}{6L} + \frac{2\pi \tilde{v}_F}{L}\,
    \tilde{X}_{0,\frac{1}{2}}
  =  -\frac{\pi\tilde{v}_F}{6L} + \frac{2\pi \tilde{v}_F}{L}\,
  \frac{\pi}{4\gamma} \,
\end{equation}
which grows as $\gamma\to0$, eventually leaving the range of applicability of
the finite size analysis based on Eq.~(\ref{fse-fm}).

\begin{figure}[t]
\includegraphics[width=0.95\textwidth]{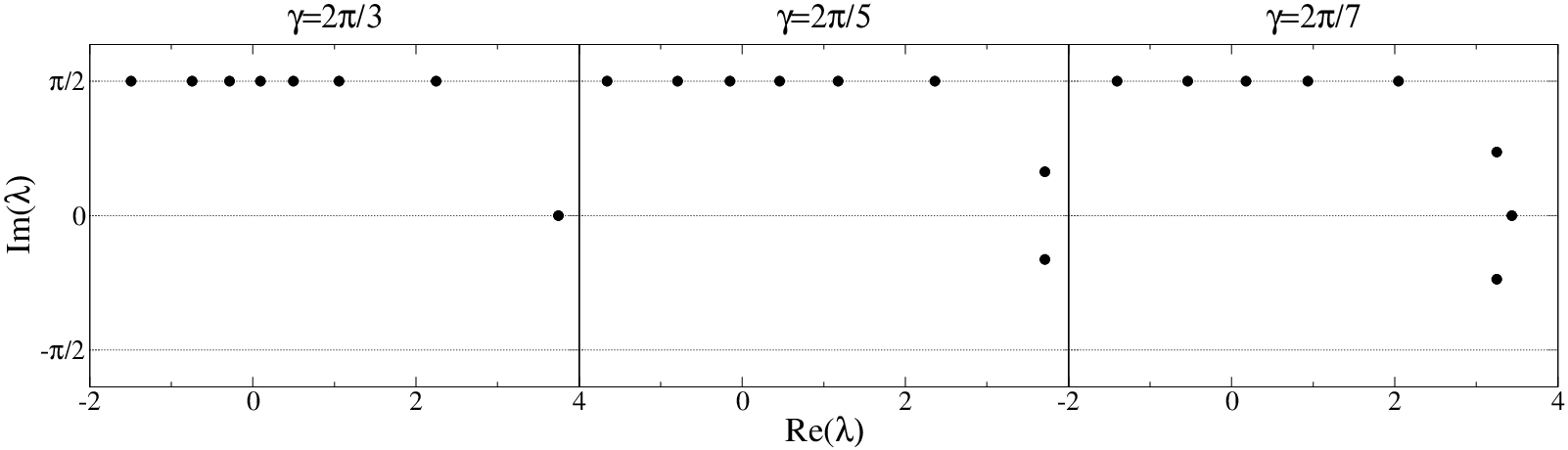}
\caption{\label{fig:xxzex} Configuration of Bethe roots for the lowest $S_3=0$
  state in the ferromagnetic disordered regime of the XXZ spin-1 chain with
  $L=8$ sites for various values of $\gamma$.}
\end{figure}

For additional insights into the properties of the spin-1 chain we have to
rely on the analysis of the Bethe equations (\ref{betheXXZ}): the
configuration of Bethe roots for this state evolves as $\gamma$ decreases from
$\pi$ to $0$: for $\gamma \in (\pi/(k+1),\pi/k)$ we find that it consists of
$L-k$ $(1,-)$-strings and a single $(k,+)$-string according to the
classification (\ref{stringXXZ}), see Fig.~\ref{fig:xxzex}.  In the finite
system the configuration reduces to a single $(L,+)$-string for
$\gamma<\pi/L$.  This is a bound state of magnon-excitations over the
ferromagnetic pseudo vacuum state with $s_3=L$.  Based on this observation we
propose that for a system of finite size $L$ there is a level crossing at
$\gamma=\pi/L$ leading to a fully polarized ground state at smaller values of
the anisotropy.  This proposal has been confirmed by numerical diagonalization
of the Hamiltonian.

\section{Antiferromagnetic regime of the mixed chain}
\label{sec:mix-afm}
To discuss the properties of the mixed superspin chain we also need to
distinguish two cases:
\begin{description}
\item[$0\le\gamma<\pi$] in analogy to the XXZ spin-1 chain we will call this
  regime 'antiferromagnetic', 
\item[$\pi<\gamma\le2\pi$] this 'ferromagnetic' regime will be discussed in
  the next section.
\end{description}

The solutions to the Bethe equations for the mixed superspin chain
(\ref{betheM}) can be classified into strings -- similarly as for the XXZ
spin-1 chain in (\ref{stringXXZ}): in the large $L$ limit Bethe roots with
$Im(\lambda_j^{(\alpha)})\ne0$ or 
$\pi/2$ have to be combined such that
their differences coincide with poles or zeroes of the bare scattering phase
shifts on the right hand sides of Eqs.~(\ref{betheM}).
As in the rational case of the $sl(2|1)$ mixed superspin chain \cite{EsFS05}
this considerations lead to 

\begin{enumerate}
\item[(1)] 1-strings\\
  just as for the XXZ spin-1 chain there are two types of unpaired roots
  allowed, namely real ones and roots on the line 
  $Im(\lambda)=\pi/2$,
\end{enumerate}
and to composites combining roots from both levels:
\begin{enumerate}
\item[(2)] wide and strange strings\\
  these configurations consist of $m$ roots from both levels of the Bethe
  ansatz with the same center $\lambda_{m}\in\mathbb{R}\cup
  \left(\mathbb{R}+i\pi/2\right)$.  For $m$ odd they have been called wide
  strings in Ref.~\onlinecite{EsFS05}, e.g.
\begin{equation}
\begin{aligned}
  \lambda^{(1)}_{m,k} &= \lambda_{m}
  +i\frac{\gamma}{4}\left(m+3-4k\right),\quad  
                     &k=1,\ldots,\frac{m+1}{2}\,,\\
  \lambda^{(2)}_{m,j} &= \lambda_{m}
  +i\frac{\gamma}{4}\left(m+1-4j\right),\quad 
                     &j=1,\ldots,\frac{m-1}{2}\,
\end{aligned}
\end{equation}
and a second type obtained by interchanging first and second level roots,
$\{\lambda^{(1)}_{m,k}\} \leftrightarrow \{\lambda^{(2)}_{m,j}\}$.  For $m$
even, the so-called strange strings have the unusual property of not being
invariant under complex conjugation, e.g.\
\begin{equation}
\begin{aligned}
  \lambda^{(1)}_{m,k} &= \lambda_{m}
  +i\frac{\gamma}{4}\left(m+3-4k\right),\quad  
                     &k=1,\ldots,\frac{m}{2}\,,\\
  \lambda^{(2)}_{m,j} &= \lambda_{m}
  +i\frac{\gamma}{4}\left(m+1-4j\right),\quad 
                     &j=1,\ldots,\frac{m}{2}\,.
\end{aligned}
\end{equation}
Again, there is a second type of such configurations for given $m$ obtained by
$\{\lambda^{(1)}_{m,k}\} \leftrightarrow \{\lambda^{(2)}_{m,j}\}$.

\item[(3)] narrow strings\\
  Finally, there are composites which contain the same number $m/2$ of roots
  on either level.  They may be seen as degenerations of two wide or strange
  strings with the same center $\lambda_m$.
\end{enumerate}
The existence of strings of a given length as
well as their parity (i.e.\ whether they are centered around the real axis or
the line 
$Im(\lambda_m)=\pi/2$) depends on the deformation parameter $\gamma$, again as
in the XXZ spin-1 chain.  For the states with lowest (and highest) energies,
we find that 1-strings of either parity and strange 2-strings centered around
the real axis plus their possible degenerations into narrow ones are
sufficient to capture the spectrum (in the 'ferromagnetic' high energy regime
this is true at least for $\gamma>\pi/3$ as discussed in
Section~\ref{sec:mix-fm} below).

In the antiferromagnetic regime this classification is particularly useful in
the sector where the total number of roots on the two levels is the same,
i.e. for $N_1=N_2$ in (\ref{betheM}): here the total $U_q[sl(2|1)]$ charge of
the Bethe state is zero.  Many low lying excited states in this sector
correspond to root configurations consisting of strange $2$-strings (up to
finite size corrections),
%
%
i.e.\ sets $\{\lambda_k^{(1)}\}$ and $\{\lambda_k^{(2)}\}$ that are mapped
onto each other by complex conjugation.
This sector also contains the lowest state in the singlet sector of the model,
where $N_1=N_2=L$.  As has been argued above, this state is also present in
spectrum of the XXZ spin-1 chain related by (\ref{ener-map}) and has energy
$E^{(mix)}\equiv 0$.  This is the ground state of the superspin chain for
$\gamma<\pi/2$.  The corresponding root configuration solving the Bethe
equations (\ref{betheM}) is $\lambda_k^{(a)}\equiv0$ for \emph{all}
$k=1,\ldots,L$ and $a=1,2$.  The same observation has been made in the
rational model obtained as $\gamma\to0$ \cite{EsFS05}.

\subsection{Small systems}

For $L$ up to $4$ we have computed the complete spectrum of the mixed chain by
exact numerical diagonalization of the Hamiltonian.  As a consequence of the
deformation some of the degenerations present in the $sl(2|1)$-symmetric
superspin chain are lifted.  For example, an $sl(2|1)$ octet $[b,s]=[0,1]$
splits into two charge 0 doublets with $s_3=\pm1$ and $s_3=0$ respectively and
a quartet with charge $\pm\frac{1}{2}$ and $s_3=\pm\frac{1}{2}$:
\begin{equation}
\begin{aligned}
  \left[0,1\right] \to &\{ |b=0,s_3=1\rangle, |b=0,s_3=-1\rangle \}\,\cup\,
            \{ |b=0,s_3=0\rangle, |b=0,s_3=0\rangle \}\,\\
            &\cup\,\{ |b=\frac{1}{2},s_3=\frac{1}{2}\rangle,
               |b=\frac{1}{2},s_3=-\frac{1}{2}\rangle,  
               |b=-\frac{1}{2},s_3=\frac{1}{2}\rangle,  
               |b=-\frac{1}{2},s_3=-\frac{1}{2}\rangle \}\,.
\end{aligned}
\end{equation}
Interestingly, we observe cases
where pairs of the $s_3=0$ doublets arising from degenerate octets in the
isotropic case split further into pairs with complex conjugate eigenvalues.

For $L=3$ we have identified the low energy states in terms of their
corresponding configuration of Bethe roots, see Table \ref{tab:L3low} for the
spectrum at anisotropy $\gamma=2\pi/7$ (the degeneracies found in the
numerical solution can be reproduced by applying symmetry operations on the
set of Bethe roots, i.e.\ $\Lambda^{(1)} \leftrightarrow \Lambda^{(2)}$ or
$\Lambda^{(a)} \leftrightarrow -\Lambda^{(a)}$ for both $a=1,2$, and by using
the global symmetries of the mixed chain, i.e.\ reversal of all spins).
\begin{table}[t,floatfix]
  \caption{\label{tab:L3low}Low energy states of the $L=3$ superspin chain in
    the antiferromagnetic regime for $\gamma=2\pi/7$ and the 
    identified Bethe configurations.  Additional root configurations to these
    energies can be obtained by using the symmetries of the Bethe equations.} 
\begin{tabular}{c|l|c|l}
\hline
$(N_1,N_2)$& Energy $E$ & degeneracy &Bethe roots\\
\hline
$(3,3)$ & $0$ & 1 & $\Lambda^{(1)} = \{0,0,0\} = \Lambda^{(2)}$ (XXZ) \\
$(2,2)$ & $1.2968$ & 2 & 
  $\Lambda^{(1)}=\{\pm i0.2386\}=\Lambda^{(2)}$ (XXZ) \\
$(3,3)$ & $1.6523$ & 2 & $\Lambda^{(1)} = \{-0.0110\pm i0.2348,\infty\}
  = -\Lambda^{(2)} $\\
$(3,2)$ & $2.3027$ & 4 & $\Lambda^{(1)} = \{ \pm i0.2492, i\pi/2 \}$, 
                         $\Lambda^{(2)} = \{ \pm i0.2255 \}$\\
$(2,2)$ & $4.9748$ & 4 &
  $\Lambda^{(1)}=\{\pm0.1161+i0.2704\}=(\Lambda^{(2)})^*$\\
$(3,3)$ & $5.1859\pm i0.6690$ & 2*2 &
  $\Lambda^{(1)} = \{-0.1429-i0.2668,\, 0.0932-i0.2675,\,\infty \}
  =-\Lambda^{(2)}$\\
%
$(3,2)$ & $6.1084\pm i0.3325$ & 2*8 &
 $\Lambda^{(1)} = \{-0.0889+i0.2822,0.1199+i0.3044,0.2581-i0.8296\}$\\
 &&& $\Lambda^{(2)} = \{-0.0790-i0.2295,0.1415-i0.2039\}$\\
\hline
\end{tabular}
\end{table}
Although they are strongly deformed in some cases, the string content of these
configurations according to the classification given above can be identified.

\begin{figure}[t,floatfix]
\includegraphics[width=0.85\textwidth]{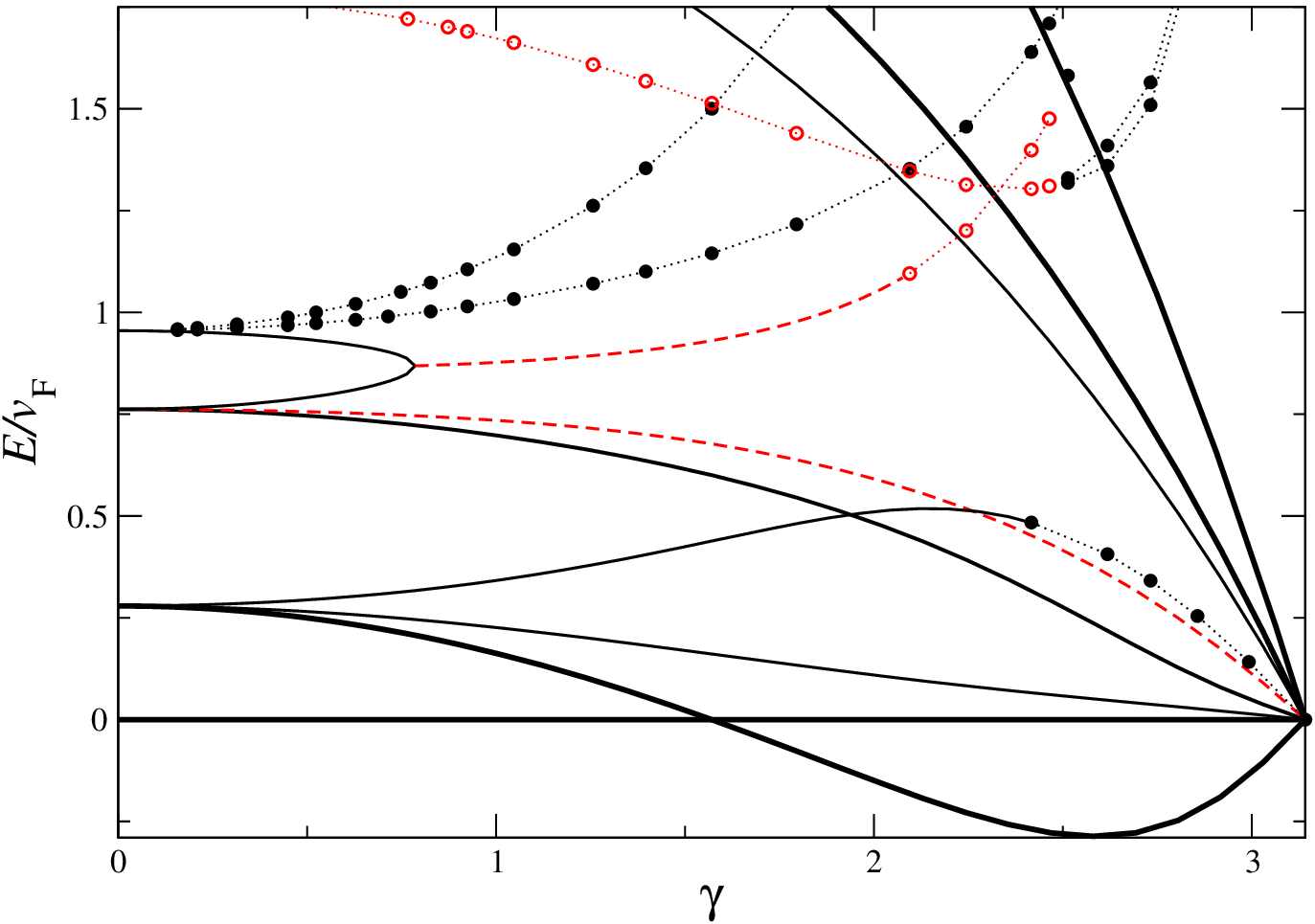}
\caption{\label{fig:L3gam} Low energy part of the spectrum of the $L=3$ chain
  for anisotropy $0\leq\gamma<\pi$: black lines and filled symbols denote the
  real eigenvalues, red lines and open symbols the real part of complex
  eigenvalues of the Hamiltonian.  Full and dashed lines denote eigenvalues
  obtained by solution of the Bethe equations, see Table~\ref{tab:L3low},
  whereas the symbols are eigenvalues obtained by numerical diagonalization of
  the Hamiltonian for which the corresponding configuration of Bethe roots has
  not been identified.  Dotted lines connecting numerical data are
  guides to the eye only.  }
\end{figure}

Based on the numerical and analytical data the following general picture for
the lowest excitations emerges:\\
As shown in Ref.~\onlinecite{EsFS05} the low-lying multiplets of the $sl(2|1)$
symmetric chain ($\gamma\to0$) are the singlet ground state with $E=0$ in the
normalization used here followed by a single $sl(2|1)$ octet as the lowest
excitation.  Above these there are two more degenerate octets and a pair of
degenerate indecomposables, each containing 8 states.
Upon deformation the degeneracies of these excitations lifted as described
above, see Fig.~\ref{fig:L3gam}.
Note that the $s_3=1$ doublet arising from the lowest octet is part of the XXZ
spin-1 subsector of the spectrum which according to Eq.~(\ref{xxzcftodd}) has
a finite size energy gap 
$\Delta E^{(XXZ)} = (2\pi v_F/L) \left[X_c-\frac{1}{8}\right]$ in the large
$L$ limit.  For $\pi/2<\gamma<\pi$ this doublet is the ground state of the
mixed chain, as expected from the analysis of the XXZ-1 chain with
anti-periodic boundary conditions for odd $L$.  As $\gamma\to\pi$ all states
evolving from this octet degenerate at $E=0$.

The 16 states of the degenerate octets split into a quartet and an octet with
real energies and two doublets with complex conjugate eigenvalues of the
Hamiltonian.  The latter eigenvalues also approach $0$ as $\gamma\to\pi$.
Similarly the states of the degenerate indecomposables split
into two quartets and an octet, all with real energies.
Increasing $\gamma$ beyond $\approx \pi/4$ the real octets from these two
groups degenerate and turn into two octets with complex conjugate energies.
Our numerical data indicate that this conversion of pairs of real eigenvalues
into pairs of complex conjugate ones appears in several regions of the
spectrum.  We have not been able to study whether this phenomenon persists as
the system size $L$ is increased, but it is a common feature in non-unitary
models.

\begin{figure}[t,floatfix]
\includegraphics[width=0.85\textwidth]{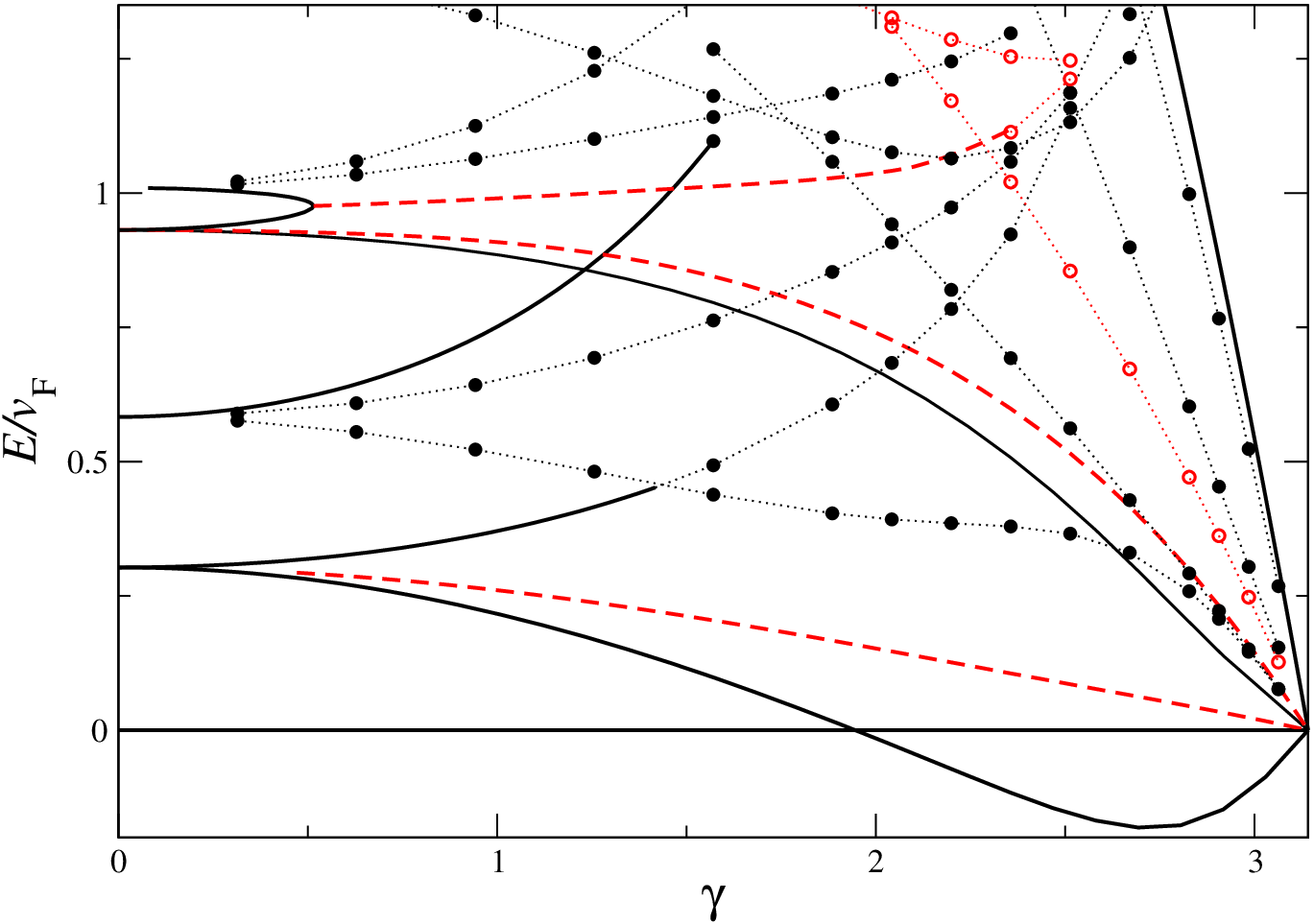}
\caption{\label{fig:L4gam} Same as Fig.~\ref{fig:L3gam} but for $L=4$.  }
\end{figure}

In Figure~\ref{fig:L4gam} we present our data for the $\gamma$ dependence of
the low energy spectrum of the $L=4$ mixed superspin chain.  
As $\gamma\to0$ the spectrum (ordered by energy) consists of the $E=0$ singlet
ground state, two degenerate octets, an 8-dimensional indecomposable and
another pair of degenerate octets \cite{EsFS05}.  The observed splittings and
appearance of levels with complex conjugate eigenvalues fit into the scheme
discussed for $L=3$ above.

\subsection{Analysis of the finite size spectrum -- antiferromagnetic regime}
Due to the identification (\ref{ener-map}) of the spectrum of the XXZ spin-1
chain within that of the superspin chain we already know part of the finite
size scaling amplitudes -- relative to the $E=0$ eigenstate -- in the latter:
\begin{equation}
\label{fse-compact}
\begin{aligned}
  \Delta E^{(mix)} = \frac{2\pi v_F}{L}\left(
    2 X^{(n,m)}_{(r,j)}(\gamma) -\frac{1}{4} \right)\,.
\end{aligned}
\end{equation}
In translating the energies from the XXZ model we have used that as a
consequence of (\ref{transf-map}) the Fermi velocity remains the same,
$v_F^{(mix)}=v_F=2\pi/\gamma$, and therefore the scaling dimensions of the
superspin chain are twice of those given in (\ref{fse-cft}) for the composite
fields in the XXZ spin-1 chain.  This is in agreement with the observations in
Ref.~\onlinecite{EsFS05}.

For $\gamma<\pi/2$ the ground state is the unique state with $E_0(L)=0$, hence
the central charge of the model is
\begin{equation}
  c = - \frac{6L}{\pi v_F}\,E_0 =  0\,,
  \qquad 0\le\gamma<\frac{\pi}{2}\,.
\end{equation}
The lowest excitation corresponds to the conformal operator with scaling
dimension $X_{(0,0)}^{(1,0)}=X_c$ in the $\pi$-twisted XXZ spin chain of odd
length $L$ (\ref{xxzcftodd}).  In the mixed chain this translates into a
scaling dimension with scaling dimension $2X_c-1/4= (\pi-2\gamma)/4\pi$
according to (\ref{fse-compact}).

\begin{figure}[t,floatfix]
\includegraphics[width=0.85\textwidth]{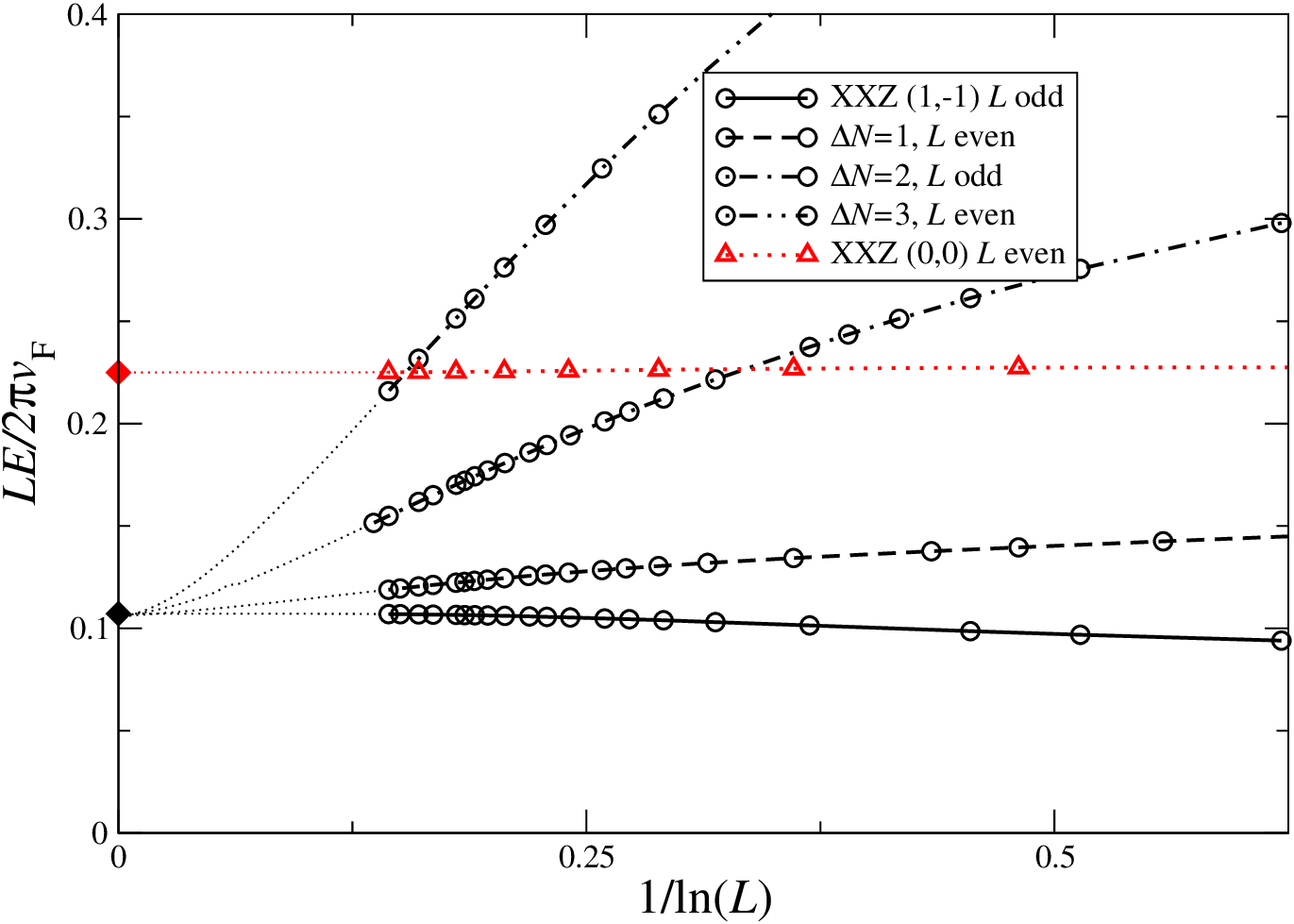}
\caption{\label{fig:fsc_g3.5} Evolution of the fine structure of the low
  energy spectrum of the mixed spin chain as a function of $L$ for
  $\gamma=2\pi/7$: circles denote energies scaling to the $(n,m)=(1,-1)$ level
  of the XXZ spin chain with odd length, triangles denote the $(0,0)$ level of
  the XXZ chain.  The dotted lines connecting to $L=\infty$ are rational
  function extrapolations of the numerical data.}
\end{figure}

In the superspin chain this state is realized by $(L-1)/2$ narrow 2-strings.
As mentioned above, a narrow string may be viewed as degeneration of two
strange 2-strings of opposite type.  Excitations can be created by lifting
this degeneracy into configurations with different number of the two possible
types of strange 2-strings, i.e.\ type '$+$'-strings with
$\lambda^{(1)}= (\lambda^{(2)})^*=\lambda+i\gamma/4$ 
and '$-$'-strings with
$\lambda^{(1)}= (\lambda^{(2)})^*= \lambda-i\gamma/4$.  The narrow string
state is described by the same number of $\pm$ strange strings, $\Delta
N=N_+-N_-=0$.  Similar as in Ref.~\onlinecite{EsFS05} one can show that states
with different but finite $\Delta N$ have the same energy in the thermodynamic
limit (note, however, that configurations with $\Delta N$ even (odd) are only
possible for $L$ odd (even)).  Computing the energies of these states we find a
logarithmic fine structure on top of the level corresponding to the operator
$(n,m)=(1,0)$, i.e.
\begin{equation}
\label{fse-nonc}
\begin{aligned}
  \frac{L}{2\pi v_F}\,\Delta E^{(mix)} = \frac{\pi-2\gamma}{4\pi} +
  K(\gamma,L)(\Delta N)^2\,,
\end{aligned}
\end{equation}
see Fig.~\ref{fig:fsc_g3.5} for $\gamma=2\pi/7$.  We have also displayed the
$L$-dependence of the $\Delta N=0$ level corresponding to the twisted spin
chain excitation with $(n,m)=(0,1)$ for even $L$ (\ref{xxzcfteven}): this
state leads to a scaling dimension $\frac{1}{4}(\pi+\gamma)/(\pi-\gamma)$ in
the spectrum of the mixed superspin chain and is part of the lowest $sl(2|1)$
indecomposable for even $L$ as $\gamma\to0$.

\begin{figure}[t]
\includegraphics[width=0.85\textwidth]{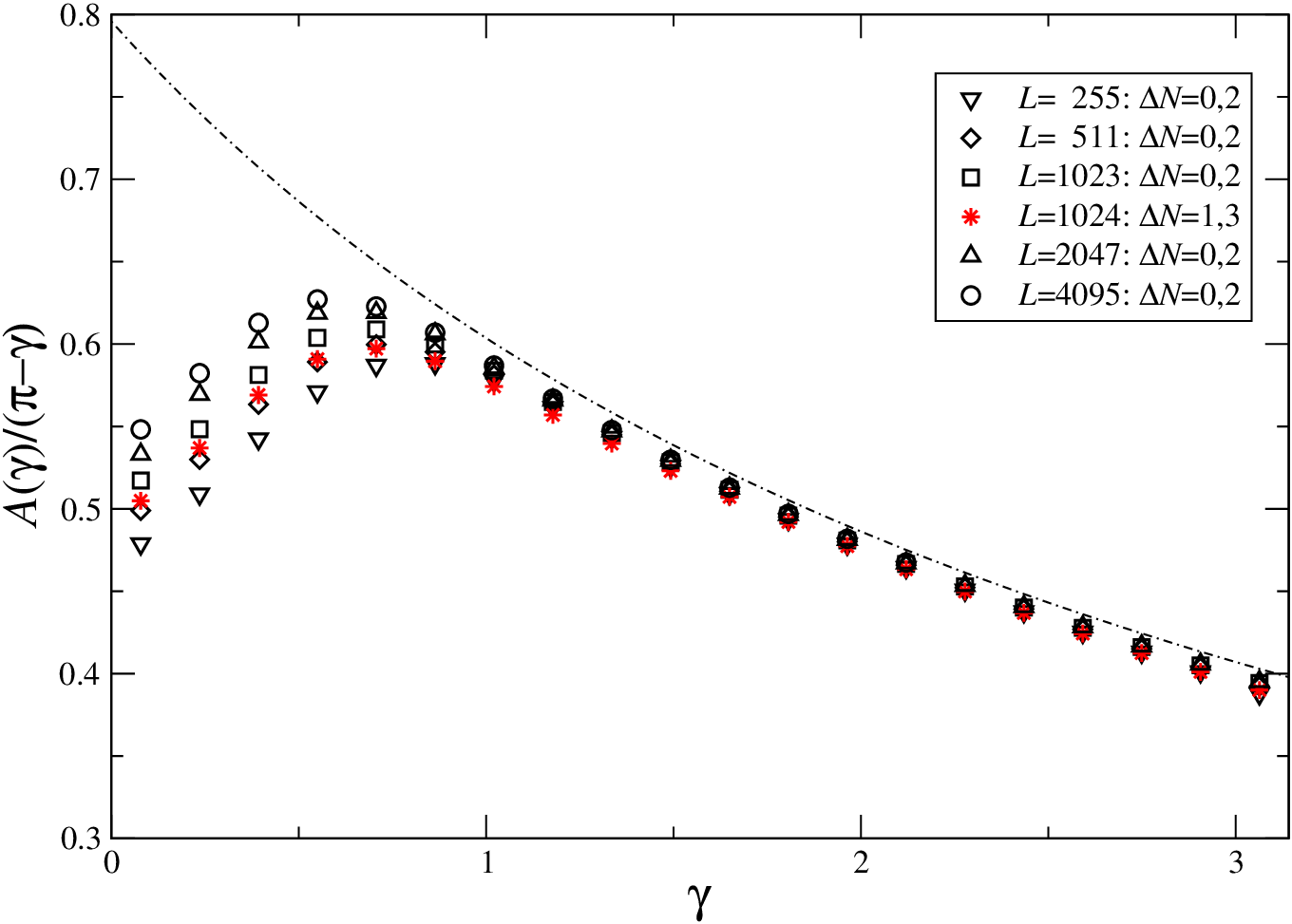}
\caption{\label{fig:fss_gam} Amplitude of the logarithmic fine structure
  extracted from comparison of the energies of corresponding states at sizes
  $L=2L'+1$ ($2L'$ for $L$ even) and $L'$.  The line is the conjectured
  $\gamma$-dependence Eq.~(\ref{fss-conj}).}
\end{figure}

Starting from a non-linear sigma model with a supersphere as target space
Ikhlef \emph{et al.} \cite{IkJS08} have proposed that this logarithmic fine
structure in the finite size spectrum is the signature of a non-compact boson
in the continuum theory.  At intermediate scales, given by the size $L$ of the
lattice regularization, the renormalization of the coupling constant leads to
an effective radius of the non-compact boson given as
\begin{equation}
  (R_{nc})^2 \propto \frac{1}{A(\gamma)}\,\left[\ln(L/L_0)\right]^2\,.
\end{equation}
This, in turn, generates the logarithmic structure (\ref{fse-nonc}) observed
in the numerical data:
\begin{equation}
\label{Ksubl}
  K(\gamma,L) = \frac{A(\gamma)}{\ln(L/L_0)^2}
\end{equation}
In Fig.~\ref{fig:fss_gam} we present numerical estimates of the amplitude
$A(\gamma)$ for the lowest excitation based on the finite size spectrum for
system sizes up to $L=4095$.
We find that the amplitude is well described by
\begin{equation}
\label{fss-conj}
  A(\gamma) = \frac{5}{2}\,\frac{\pi-\gamma}{\pi+\gamma}\,.
\end{equation}
For $\gamma\gtrsim 1$ the extrapolation of the numerical data is within 1\% of
this conjecture.  For more convincing evidence, in particular at small values
of the deformation parameter $\gamma$, one would need additional information
about higher order corrections to (\ref{Ksubl}) and have to study system sizes
which are out of reach for this method based on the numerical solution of the
Bethe equations (\ref{betheM}).

\begin{figure}[t]
\includegraphics[width=0.85\textwidth]{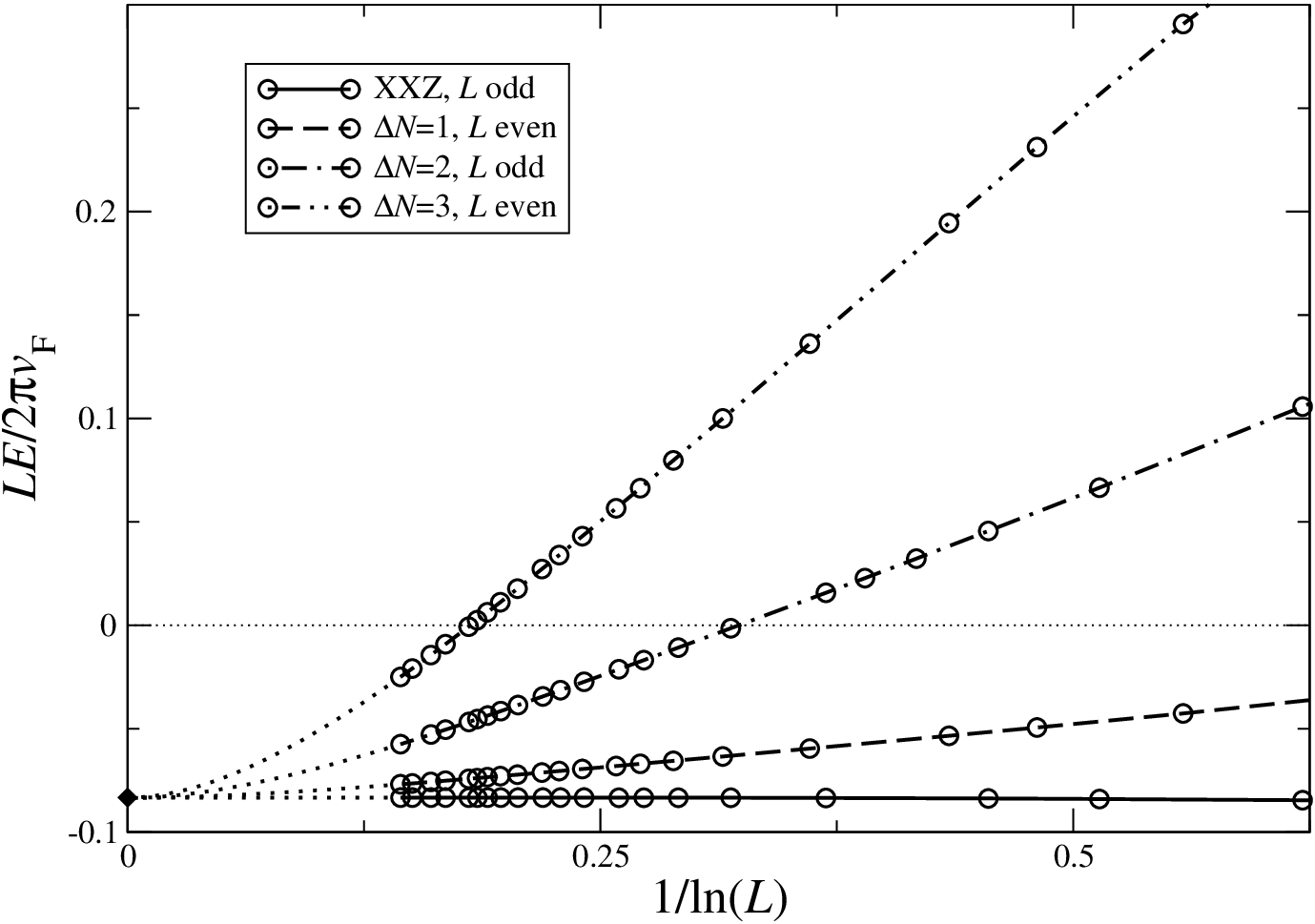}
\caption{\label{fig:fsc_g1.5} As Fig.~\ref{fig:fsc_g3.5} but for
  $\gamma=2\pi/3$.  The dotted line denotes the energy level at $E=0$ of the
  singlet state without any finite size scaling in the spectrum of the
  mixed chain for $L$ both even and odd.}
\end{figure}

For $\pi/2<\gamma<\pi$ the set of levels (\ref{fse-nonc}) has energies below
the $E\equiv0$ eigenstate of the mixed chain for $L$ sufficiently large, see
Fig.~\ref{fig:fsc_g1.5}.  
As a consequence the model is in a different
universality class with central charge determined by the finite size scaling
behaviour of the lowest level in this set, i.e.\
\begin{equation}
  c_\mathrm{eff}(\gamma) = - \frac{6L}{\pi v_F}\,E^{(mix)} =
  3\,\frac{2\gamma-\pi}{\pi}\,, 
  \qquad \frac{\pi}{2}<\gamma<\pi\,.
\end{equation}
Note that this state can be identified with the lowest level of the XXZ spin-1
chain for odd $L$ only.  For even $L$ its realization in the mixed superspin
is in terms of a strange string configuration with $\Delta N=\pm 1$.
Immediately above this ground state there is a continuum of states with energy
gaps vanishing as $\propto (\Delta N)^2/\left[L\,\ln(L/L_0)^2\right]$ due to
the presence of the non-compact boson: our finite size analysis indicates that
these gaps show the same $\gamma$ dependence as in (\ref{fss-conj}) given
above.  The first excitation from the XXZ spin-1 subset of the spectrum above
this continuum is the $E\equiv0$ state corresponding to a scaling dimension
\begin{equation}
  X = \frac{2\gamma-\pi}{4\pi}\,.
\end{equation}

\section{Ferromagnetic regime of the mixed chain}
\label{sec:mix-fm}
Again we use the spectral relation (\ref{relation}) to discuss the properties
of the mixed superspin chain in this regime in the interval $0\le\gamma<\pi$ by
changing the sign of the energy eigenvalues (\ref{ener-sl}).

As for the antiferromagnetic regime above we begin our analysis  based on
numerical data for the spectrum of the $L=3$ chain obtained by exact
diagonalization of the superspin Hamiltonian, see Table~\ref{tab:L3high} for
$\gamma=2\pi/7<\pi/3$:
according to our discussion of the disordered ferromagnetic regime of the XXZ
spin-1 chain above the lowest state from the XXZ spin-1 part of the spectrum
for this value of $\gamma$ is the completely polarized reference state with
$s_3=3$ and energy $E=-12\cot(\gamma/2)$ according to (\ref{ener-map}).  In
Table~\ref{tab:L3high} we have also listed the level corresponding to the
primary field $(n,m)=(0,\frac{1}{2})$ in the twisted XXZ model (\ref{fse-fm}).

The numerical data reveal that below the reference state of the XXZ spin chain
there are many states with lower energies in the spectrum of the mixed
superspin chain in the charge-sectors $b=0,\pm1/2$ and with magnetization
$0\le s_3\le L$: the corresponding Bethe configurations consist of $(N_1,N_2)$
roots with $N_1-N_2=0,\pm1$ and $Im(\lambda_j^{(a)})=\pi/2$, i.e.\
$(1,-)$-strings.  The ground state in the sector $(L,L-1)$ (together with the
equivalent sector $(L-1,L)$) is four-fold degenerate.

\begin{table}[t,floatfix]
  \caption{\label{tab:L3high}Low energy states of the $L=3$ superspin chain in
    the ferromagnetic regime for $\gamma=2\pi/7$ and the identified Bethe
    configurations.} 
\begin{tabular}{c|l|c|l}
\hline
$(N_1,N_2)$&Energy $E$ & degeneracy &Bethe roots\\
\hline
$(3,2)$ & $-27.8683$ & 4 & 
  $\Lambda^{(1)} = \{\pm1.1935+i\pi/2,\,i\pi/2\}$,
  $\Lambda^{(2)} =\{\pm0.4797+i\pi/2\}$\\
$(3,3)$ & $-27.7508$ & 2 &
  $\Lambda^{(1)} =\{-0.7367+i\pi/2,\,0.2722+i\pi/2,\,\infty\} =
  -\Lambda^{(2)}$\\
%
$(2,2)$ & $-27.4205$ & 4 &
  $\Lambda^{(1)} = \{-1.0384+i\pi/2,\,0.2015+i\pi/2\} =
   -\Lambda^{(2)}$\\
%
$(2,1)$ & $-27.1935$ & 4 & 
  $\Lambda^{(1)} = \{\pm0.5829 +i\pi/2\},\,
   \Lambda^{(2)} = \{i\pi/2\}$\\
%
$(1,1)$ & $-26.4469$ & 4 &
  $\Lambda^{(1)} =\{-0.4447+i\pi/2\} = -\Lambda^{(2)}$\\
%
%
$(1,0)$ & $-25.8814$ & 4 & $\Lambda^{(1)} = \{i\pi/2\}$\\
%
$(0,0)$ & $-24.9183$ & 2 & (XXZ pseudo vacuum) \\
%
$(3,2)$ & $-24.8425$ & 8 &
  $\Lambda^{(1)} = \{-1.0326+i\pi/2,\, 0.3894+i\pi/2,\,
                    0.6432 \}$\\
 & && $\Lambda^{(2)} = \{-0.2736+i\pi/2,\, 0.7440\}$\\
%
$(3,3)$ & $-24.7595$ & 4 & 
  $\Lambda^{(1)} = \{-0.5477+i\pi/2,1.0369,\infty\}$\\
 & &&$\Lambda^{(2)} = \{-\infty,\,-0.0189+i\pi/2,\,0.5909\}$\\
%
$(2,2)$ & $-24.6243$ & 8 & 
  $\Lambda^{(1)} =\{-0.8222,\,-0.2226+i\pi/2\}$\\
 & &&$\Lambda^{(2)} =\{-0.7559,\,0.8563+i\pi/2\}$\\
 & \hspace{\fill}\vdots\hspace{\fill} &&\\
$(3,3)$ & $-23.6547$ & 2 &
  $\Lambda^{(1)} =\{0.6023\pm i0.6800,0.7226 \} = \Lambda^{(2)}$ (XXZ)\\
\hline
\end{tabular}
\end{table}

\subsection{Thermodynamic limit}
Based on this observation we shall now study this ground state in the
thermodynamic limit $L\to\infty$.
To make further progress it is convenient to rewrite the rapidities
$\lambda_j^{(a)}$ as,
\begin{equation}\label{newvar}
\lambda_j^{(a)}=\mu_j^{(a)} +i \frac{\pi}{2},
\end{equation}
where $\mu_j^{(a)} \in \mathbb{R} $ for $a=1,2$.
Now, by substituting Eq.~(\ref{newvar}) in the Bethe ansatz equations
(\ref{betheM}) and by taking their logarithms we find that the resulting
relations for $\mu_j^{(k)}$ are,
\begin{equation}
\label{betheMM}
\begin{aligned}
   L \Phi(\mu_j^{(1)},\gamma-\pi) &= 2\pi Q_j^{(1)} + \sum_{k=1}^{N_2}
      \Phi(\mu_j^{(1)}-\mu_k^{(2)},\gamma)\,, \qquad j=1,\ldots,N_1\\
   L \Phi(\mu_j^{(2)},\gamma-\pi) &= 2\pi Q_j^{(2)} + \sum_{k=1}^{N_1}
      \Phi(\mu_j^{(2)}-\mu_k^{(1)},\gamma)\,, \qquad j=1,\ldots,N_2
\end{aligned}
\end{equation}
where $\Phi(x,\gamma) = 2\arctan\left( \tanh(x) \cot(\gamma/2)\right)$.
%
%
The numbers $Q_j^{(a)}$ define the many possible branches of the logarithm.
They have to be chosen integer or half-odd integer depending on the parities
of $N_{a}$ according to the rule
\begin{equation}
\label{Qparity}
  Q_j^{(1)} \equiv \frac{N_2}{2} \mod 1\,,\qquad
  Q_j^{(2)} \equiv \frac{N_1}{2} \mod 1\,.
\end{equation}
For example the ground state in the sector $N_1=L$ and $N_2=L-1$ is described
by the symmetric sequences
\begin{equation}
\label{Qs}
\begin{aligned}
Q_{j}^{(1)}&= \frac{L+1}{2} -j,\qquad j=1,\dots, L,  \\
Q_{j}^{(2)}&= \frac{L}{2} -j,\qquad
j=1,\dots ,L - 1.
\end{aligned}
\end{equation}

At this point we have the basic ingredients to compute some of the
thermodynamic limit properties.  When $L \rightarrow \infty$ the number of
roots $\mu_j^{(a)}$ tend towards a continuous distribution on the real axis
whose density which we shall denote by $\rho^{(a)}(\mu)$.  The Bethe equations
(\ref{betheMM}) turn into coupled linear integral relations for the densities
$\rho^{(a)}(\mu)$ which can be solved by Fourier transform.  The fact that
Eqs.~(\ref{betheMM}) are symmetric under the exchange of rapidities
$\mu_j^{(1)} \leftrightarrow \mu_j^{(2)}$ in the thermodynamic limit implies
that $\rho^{(1)}(\mu)\equiv \rho^{(2)}(\mu)$.  The final result for such density
is,
%
\begin{equation}
\label{rhostherm}
\rho^{(a)}(\mu)= \frac{1}{(\pi-{\gamma}/2)} 
\frac{\cos\left[\frac{\pi {\gamma}}{4(\pi-{\gamma}/2)}\right] 
\cosh\left[\frac{\pi \mu}{\pi-{\gamma}/2}\right]} 
{\cosh\left[\frac{2 \pi \mu}{\pi-{\gamma}/2}\right] +
\cos\left[\frac{\pi {\gamma}}{2(\pi-{\gamma}/2)}\right]},
\qquad\mathrm{for}~~a=1,2. 
\end{equation}
Now from the expressions for the density $\rho^{(a)}(\mu)$ and
Eq.~(\ref{ener-sl}) we can compute the ground state energy density
$\tilde{e}_\infty=E_0/L$.  By writing the infinite volume limit of
Eq.~(\ref{ener-sl}) in terms of its Fourier transform we find
\begin{equation}
\label{einf}
\tilde{e}_{\infty} = -4 \cot({\gamma}/2) -4 \int_{0}^{\infty} 
\mathrm{d} \omega \frac{
\sinh[\omega {\gamma}/2] \cosh[\omega {\gamma}/4]}{\sinh[\omega \pi/2] 
\cosh[\omega(2\pi-{\gamma})/4]}
\qquad\mathrm{for}\quad 0 \leq {\gamma} < \pi\,.
\end{equation}

In addition, we have verified that the low-lying excited states about the
ground state are gapless. As usual, these states can be obtained by inserting
holes in the density distribution of $\mu_j^{(a)}$ by making alternative
choices for $Q_j^{(a)}$. This procedure is nowadays familiar to many
integrable models solved by Bethe ansatz and for technical details see for
example (\cite{Suth75,FaTa81,HUBBARD}).  We find that the low-momenta dispersion
relation among the energy $\epsilon^{(a)}(\mu)$ and momenta $p^{(a)}(\mu)$,
both measured from the ground state, has a relativistic behaviour
\begin{equation}
\epsilon^{(a)}(\mu) \sim \tilde{v}_F^{(mix)} p^{(a)}(\mu)
\end{equation}
The common slope at $p^{(a)}(\mu)=0$ is the corresponding Fermi velocity of
the excitations.  It is determined by
\begin{equation}
  \tilde{v}_F^{(mix)}= \left.
     \frac{\dot{\epsilon}^{(a)}(\mu)}{2\pi \rho^{(a)}(\mu)} 
  \right|_{\mu=\infty}= \frac{2 \pi}{2\pi-{\gamma}} \,.
\end{equation}
As expected from (\ref{transf-map}) it coincides with the velocity
$\tilde{v}_F$ of low energy excitations of the XXZ spin-1 chain in the
disordered ferromagnetic regime, see Section~\ref{sec:xxz-fm}.

\subsection{Analysis of the finite size spectrum -- ferromagnetic regime}
From our investigation of the behaviour of the Bethe ansatz roots associated
to the low-lying excitations we found that they can be well described in terms
of real rapidities.  The roots with fixed imaginary part at $i\frac{\pi}{2}$
can easily be mapped on real roots by means the straightfoward shift
(\ref{newvar}).  We remark however that some excitations have the peculiar
feature that some of the their roots have the real part located at infinity.
This scenario suggests us that a first insight on the structure of the
finite-size corrections can be obtained by applying the standard density root
method, see for instance \cite{VeWo85,WoEc87b,Vega88,Suzu88,HUBBARD}.  This
technique explores the Bethe ansatz solution and it allows to compute the
$O\left( L^{-2} \right)$ corrections to the densities of roots
$\rho^{(k)}(\mu)$.  This approach predicts that the finite-size corrections to
the low-lying energies eigenvalues have the following form,
\begin{equation}
\label{FSC}
E(L,{\gamma})  -L \tilde{e}_{\infty} =
  \frac{2\pi \tilde{v}_{F}}{L} \left[ -\frac{1}{6} +
  X_{n_{1},n_{2}}^{m_1,m_{2}}({\gamma}) \right] 
+ o\left( L^{-1} \right),
\end{equation}
where the scaling dimensions $X_{n_{1},n_{2}}^{m_1,m_{2}}({\gamma})$
depend on the anisotropy ${\gamma}$ as
\begin{equation}
\label{dimens}
\begin{aligned}
  X_{n_{1},n_{2}}^{m_1,m_{2}}({\gamma}) =& 
      \frac{1}{4} \left(1-\frac{\gamma}{2\pi}\right) ( n_1-n_2)^2
    + \frac{1}{4} \left(1-\frac{\gamma}{2\pi}\right)^{-1} ( m_1-m_2)^2\\
    &+ \frac{1}{4} \left(\frac{\gamma}{2\pi}\right) ( n_1+n_2)^2
    + \frac{1}{4} \left(\frac{2\pi}{\gamma}\right) ( m_1+m_2)^2\,.
\end{aligned}
\end{equation}
In (\ref{dimens}) the integers $n_1$ and $n_2$ are related to the number of
roots at each level of the Bethe equations by $N_1=L-n_1$ and $N_2=L-n_2$.
Therefore $(n_1\pm n_2)/2$ are associated to the conserved $U(1)$ spin $s_3$
and charge $b$ of the model: the scaling dimensions (\ref{dimens}) exhibit
exact spin charge separation in the low energy effective theory.  The
corresponding excitations of the model are free bosons with compactification
radii $R_s^2\sim\gamma/2\pi$ and $R_h^2\sim (1-\gamma/2\pi)$, usually
denominated spinon and holon modes.\footnote{%
  A similar observation has been made in one of the critical phases of a
  Temperley Lieb model with staggered spectral parameters where the effective
  field theory consists of a compact boson and two Majorana fermions
  \cite{IkJS09}.  In this model, however, the low energy degrees of freedom
  cannot be related to $U(1)$ charges of the microscopic model.}
The indices $m_1$ ($m_2$) are related to macroscopic momentum of the
excitation due to backscattering processes on the first (second) level of the
Bethe ansatz and they are usually called vortex excitations.  As a consequence
of (\ref{Qparity}) they take integer (half-odd integer) values depending on
the parity of $N_1+N_2$ leading to the following constraint connecting spinon
and vortex numbers:
\begin{equation}
\label{const3}
\begin{aligned}
  &\bullet \quad \mathrm{for~} n_1\pm n_2 \mathrm{~odd~}
    &\rightarrow \qquad & m_1,m_{2} = 0, \pm 1 ,\pm 2, \dots\\
  &\bullet \quad \mathrm{for~} n_1\pm n_2 \mathrm{~even~}\quad
    &\rightarrow \qquad &m_1,m_{2} = \pm \frac{1}{2}, \pm \frac{3}{2} 
            ,\pm \frac{5}{2} , \dots\,\, .
\end{aligned}
\end{equation}

Let us now investigate the validity of the formulae (\ref{dimens}) for the
conformal dimensions together with the selection rule (\ref{const3}).  In
order to do that we have solved the Bethe ansatz equations for a number of
low-lying states up to $L=32$.  From the numerical data we compute the
sequence
\begin{equation}
\label{FSSestim}
X(L) = \frac{L}{2\pi \tilde{v}_F}\left( E(L,{\gamma}) -
  L\tilde{e}_{\infty} \right)  + \frac{1}{6}
\end{equation}
which in the thermodynamic limit is expected to extrapolate to the dimensions
(\ref{dimens}). 

In Table~\ref{one} we show the finite-size sequences (\ref{FSSestim}) for
the ground state $E_0(L,{\gamma})$ in the case of various values of
${\gamma}$. 
\begin{table}
\caption{\label{one}Finite size sequences \ref{FSSestim} of the anomalous
  dimension  $X_{1,0}^{0,0}({\gamma})$    for ${\gamma} = \pi/6$, $\pi/3$,
  $\pi/2$, $2\pi/3$ from the Bethe   ansatz.  The expected exact conformal
  dimension is  $X_{1,0}^{0,0}({\gamma})=\frac{1}{4}$. 
}
\begin{tabular}{|c||c|c|c|c|}
  \hline
$X_{1,0}^{0,0}({\gamma})$ & $\frac{\pi}{6}$ & $\frac{\pi}{3}$ & $\frac{\pi}{2}$ & $ \frac{2\pi}{3}$ \\ \hline \hline
  4      & 0.20487467 & 0.23017601 & 0.24810873 & 0.24554137   \\
  8     & 0.23375071 & 0.24621304 & 0.24978827 & 0.24940359   \\
  12     & 0.24319443 & 0.24843089 & 0.24992216 & 0.24976865   \\
  16     & 0.24643218 & 0.24913601 & 0.24995927 & 0.24987579   \\
  20     & 0.24780188 & 0.24945223 & 0.24997482 & 0.24992223   \\
  24     & 0.24850243 & 0.24962151 & 0.24998285 & 0.24994660   \\
  28     & 0.24891117 & 0.24972275 & 0.24998754 & 0.24996104   \\
  32     & 0.24917161 & 0.24978814 & 0.24999053 & 0.24997031 \\ \hline
  Extrap.& 0.2503(2)  & 0.250003(1)& 0.250002(2)& 0.250001(2)    \\ \hline
  Exact  & 0.25  & 0.25 & 0.25  & 0.25  \\ \hline
\end{tabular}
\end{table}
The extrapolated value of the corresponding conformal dimension turns out to
be independent of the anisotropy ${\gamma}$ whose value is in good accordance
with the one predicted by Eq.~(\ref{dimens}) for $X_{1,0}^{0,0}(\gamma)=
X_{0,1}^{0,0}(\gamma)\equiv \frac{1}{4}$.  
From this result we find that the ground state energy scales as
\begin{equation}
\label{FSCgs}
  E_0(L,{\gamma}) - L \tilde{e}_{\infty} = \frac{\pi v_F}{6L}
  + o\left( L^{-1} \right),
\end{equation}
which leads us to conclude that the continuum limit of the superspin chain in
the disordered ferromagnetic regime should be described by a conformally
invariant theory with central charge $c=-1$.  The respective anomalous
dimensions $\bar{X}_{n_1,n_2}^{m_1,m_2}({\gamma})$ of the theory have to be
measured from the ground state (\ref{FSCgs}) and therefore they should be
given by,
\begin{equation}
  \bar{X}_{n_1,n_2}^{m_1,m_2}({\gamma}) = 
    X_{n_1,n_2}^{m_1,m_2}({\gamma}) -\frac{1}{4}\,.  
\end{equation}
We have also analyzed the corrections to scaling due to the presence of
irrelevant operators in the lattice Hamiltonian \cite{Card86a}.  For all
values of $\gamma$ we found the leading corrections to the finite size
estimate (\ref{FSSestim}) for $X_{0,1}^{0,0}(\gamma)$ to be of order $L^{-2}$
arising from the conformal block of the identity operator which explains the
good convergence of the extrapolation.

We now turn our attention to the excited states in order to bring extra
support to the proposal (\ref{dimens}), (\ref{const3}).  The first excitation
above the ground state (\ref{FSCgs}) occurs in the sector $n_1=n_2=0$.  Our
analysis of the $L=3$ system (see Table~\ref{tab:L3high}) indicates that among
the corresponding Bethe roots there is a pair of rapidities
$(\lambda^{(1)},\lambda^{(2)})$ which takes values $(\pm\infty,\mp\infty)$.
The presence of such infinities leads to an effective scattering phase shift
of $\exp[\pm i{\gamma}]$ for the remaining roots.  Due to this peculiarity we
shall present explicitly the form of the Bethe equations for the finite roots
\begin{equation}
\label{betheMM1}
\begin{aligned}
   L \Phi(\mu_j^{(1)},\gamma-\pi) &= 2\pi Q_j^{(1)} + \gamma + \sum_{k=1}^{L-2}
      \Phi(\mu_j^{(1)}-\mu_k^{(2)},\gamma)\,, \qquad j=1,\ldots,L-1\\
   L \Phi(\mu_j^{(2)},\gamma-\pi) &= 2\pi Q_j^{(2)} -\gamma + \sum_{k=1}^{L-1}
      \Phi(\mu_j^{(2)}-\mu_k^{(1)},\gamma)\,, \qquad j=1,\ldots,L-1\,.
\end{aligned}
\end{equation}
For this state the numbers $Q_{j}^{(a)}$  are given  by
\begin{equation}
\label{Qs1}
\begin{aligned}
  Q_{j}^{(1)}&= \frac{L+1}{2} -j,\qquad  j=1,\dots, L-1\,,  \\
  Q_{j}^{(2)}&= -\frac{L+1}{2} +j,\qquad  j=1,\dots ,L - 1\,.
\end{aligned}
\end{equation}
In table (\ref{two}) we present the finite-size estimates associated to the
state described by Eqs.~(\ref{betheMM1}) with (\ref{Qs1}).  We observe that
the the extrapolated values agree with the proposal (\ref{dimens}),
(\ref{const3}) which predicts that the lowest conformal dimension is
$\bar{X}_{0,0}^{\half,-\half}({\gamma}) = \frac{{\gamma}}{4(2\pi-{\gamma})}$.
\begin{table}
\caption{\label{two}Finite size sequences \ref{FSSestim} of the anomalous
  dimension 
  $X_{0,0}^{\half,-\half}({\gamma})$ for ${\gamma} = \pi/6$, $\pi/3$, $\pi/2$,
  $2\pi/3$ from the Bethe ansatz. The expected exact conformal
  dimension is $X_{0,0}^{\half,-\half}(\gamma)= \pi/(4 \pi-2\gamma)$.
}
\begin{tabular}{|c||c|c|c|c|}
  \hline
$X_{0,0}^{\half,-\half}({\gamma})$ & $\frac{\pi}{6}$ & $\frac{\pi}{3}$ 
       & $\frac{\pi}{2}$ & $ \frac{2\pi}{3}$ \\ \hline \hline
   4     & 0.23109569 & 0.28352514 & 0.33170562 & 0.36094436   \\ 
  8     & 0.25745093 & 0.29695955 & 0.33325340 & 0.37254539   \\ 
  12     & 0.26633674 & 0.29875319 & 0.33331810 & 0.37397623   \\ 
  16     & 0.26938340 & 0.29931576 & 0.33332857 & 0.37443588   \\ 
  20     & 0.27066931 & 0.29956684 & 0.33333139 & 0.37464230   \\ 
  24     & 0.27132594 & 0.29970094 & 0.33333245 & 0.37475283   \\ 
  28     & 0.27170870 & 0.29978104 & 0.33333283 & 0.37481895   \\ 
  32     & 0.27211767 & 0.29983274 & 0.33333303 & 0.37486165 \\ \hline 
 Extrap.&  0.27273(2) & 0.300004(2)& 0.3333332(2)& 0.375001(2)  \\ \hline 
  Exact  & 0.272727 $\dots$  & 0.3 & 0.333333 $\cdots$  & 0.375  \\ \hline
\end{tabular}
\end{table}

In Tables \ref{three}-\ref{six} we present the finite-size sequences for
several  low-lying excitations in other sectors.  We observe that the
extrapolated values corroborate the result predicted by (\ref{dimens}). 
\begin{table}
\caption{\label{three}
  Finite size sequences \ref{FSSestim} of the anomalous dimension
  $X_{1,1}^{\half,-\half}({\gamma})$  for ${\gamma} = \pi/6$, $\pi/3$,
  $\pi/2$, $2\pi/3 $ from the Bethe ansatz. The expected exact conformal
  dimension is $X_{1,1}^{\half,-\half}({\gamma})=
  \frac{{\gamma}}{2\pi}+\frac{\pi}{2(2\pi-{\gamma})}$.  
}
\begin{tabular}{|c||c|c|c|c|}
  \hline
$X_{1,1}^{\half,-\half}({\gamma})$ & $\frac{\pi}{6}$ & $\frac{\pi}{3}$ & $\frac{\pi}{2}$ & $ \frac{2\pi}{3}$ \\ \hline \hline
  4     & 0.31054918 & 0.44669719 & 0.58683657 & 0.74133601   \\ 
  8    & 0.33974964 & 0.46283122 & 0.58433551 & 0.71638581   \\ 
  12     & 0.34923933 & 0.46507338 & 0.58378663 & 0.71189057   \\ 
  16     & 0.35248317 & 0.46578873 & 0.58358979 & 0.71032924   \\ 
  20     & 0.35385561 & 0.46610987 & 0.58349789 & 0.70960905   \\ 
  24     & 0.35455798 & 0.46628188 & 0.58344777 & 0.70921857   \\ 
  28     & 0.35496794 & 0.46638478 & 0.58341748 & 0.70898230   \\ 
  32     & 0.35522923 & 0.46645125 & 0.58339779 & 0.70883087 \\ \hline 
 Extrap. & 0.35603(2) & 0.466665(2) & 0.5833334(1) & 0.708332(2)  \\ \hline
  Exact  & 0.356060 $\dots$  & 0.466666 $\cdots$ & 0.583333 $\cdots$  & 0.708333 $\cdots$ \\ \hline
\end{tabular}
\end{table}
\begin{table}
\caption{\label{four} Finite size sequences \ref{FSSestim} of the anomalous
  dimension $X_{1,2}^{0,0}({\gamma})$ for ${\gamma} = \pi/6$, $\pi/3$,
  $\pi/2$, $2\pi/3$ from the Bethe ansatz. The expected exact conformal
  dimension is $X_{1,2}^{0,0}({\gamma})=\frac{1}{4} +{\gamma}/\pi$ 
}
\begin{tabular}{|c||c|c|c|c|}
  \hline
$X_{1,2}^{0,0}({\gamma})$ & $\frac{\pi}{6}$ & $\frac{\pi}{3}$ & $\frac{\pi}{2}$ & $ \frac{2\pi}{3}$ \\ \hline \hline
  4     &0.36660004  & 0.56006032 & 0.75496874 & 0.95242531   \\ 
  8     &0.39901997  & 0.57907838 & 0.75200345 & 0.92788809  \\ 
  12     &0.40929983  & 0.58158419 & 0.75094335 & 0.92183973   \\
  16     &0.41280124  & 0.58237304 & 0.75054074 & 0.91961495   \\
  20     &0.41428312  & 0.58272535 & 0.75034904 & 0.91856536   \\
  24     &0.41504200  & 0.58291356 & 0.75024350 & 0.91798979   \\
  28     &0.41548512  & 0.58302599 & 0.75017939 & 0.91764084   \\
  32     &0.41576759 & 0.583098552 & 0.75013758 & 0.91741358 \\ \hline 
  Extrap.& 0.4166(2) &0.583332(1) & 0.750001(2) & 0.916662(2)   \\ \hline
  Exact  & 0.41666 $\cdots$  & 0.58333 $\cdots$ & 0.75  & 0.91666 $\cdots$  \\ \hline
\end{tabular}
\end{table}
\begin{table}
\caption{\label{five}Finite size sequences \ref{FSSestim} of the anomalous
  dimension $X_{1,0}^{1,-1}({\gamma})$  for ${\gamma} = \pi/6$, $\pi/3$,
  $\pi/2$, $2\pi/3 $ from the Bethe ansatz. The expected exact conformal
  dimension is
  $X_{1,0}^{1,-1}({\gamma})=\frac{1}{4}+\frac{2\pi}{(2\pi-{\gamma})}$. 
}
\begin{tabular}{|c||c|c|c|c|}
  \hline
$X_{1,0}^{1,-1}({\gamma})$ & $\frac{\pi}{6}$ & $\frac{\pi}{3}$ & $\frac{\pi}{2}$ & $ \frac{2\pi}{3}$ \\ \hline \hline
  4      & 2.78965545 & 1.76360998 & 1.5577620 & 1.59077025 \\ 
  8     & 1.99919576 & 1.58255107 & 1.5863263 & 1.70990754 \\ 
  12     & 1.69307777 & 1.51547069 & 1.5856105 & 1.73214654  \\
  16     & 1.55596049 & 1.48826402 & 1.5848095 & 1.73994982 \\ 
  20     & 1.48454418 & 1.47493397 & 1.5843369 & 1.74356556 \\ 
  24     & 1.44315270 & 1.46748697 & 1.5840527 & 1.74553075 \\ 
  28     & 1.41720088 & 1.46292463 & 1.5838718 & 1.74671607 \\ 
  32     & 1.39992651 & 1.45993403 & 1.5837506 &  1.7474855 \\ \hline
Extrap.  & 1.3404(3) &  1.45003(1) & 1.5834(2) & 1.75002(1)   \\ \hline
  Exact  & 1.340909 $\cdots$  & 1.45  & 1.58333 $\cdots$  & 1.75  \\ \hline
\end{tabular}
\end{table}
\begin{table}
\caption{\label{six}Finite size sequences \ref{FSSestim} of the anomalous
  dimension  $X_{0,0}^{\half,\half}({\gamma})$  for ${\gamma} = 2\pi/7$,
  $2\pi/5$,   $2\pi/3$, $5\pi/6$ from the Bethe ansatz. The expected exact
  conformal   dimension is  $X_{0,0}^{\half,\half}({\gamma})=\frac{\pi}{2
    {\gamma}}$.  
}
\begin{tabular}{|c||c|c|c|c|}
  \hline
$X_{0,0}^{\half,\half}({\gamma})$ & $\frac{2\pi}{7}$ & $\frac{2\pi}{5}$ & $\frac{2\pi}{3}$ & $ \frac{5\pi}{6}$ \\ \hline \hline
  4      & 1.84589078 & 1.29492470 & 0.75018486 & 0.54556892   \\ 
  8     & 1.76868129 & 1.25924483 & 0.75002911 & 0.58757296  \\ 
  12     & 1.75782587 & 1.25398066 & 0.75001177 & 0.59492733   \\
  16     & 1.75432209 & 1.25248723 & 0.75000640 & 0.59722733   \\
  20     & 1.75274406 & 1.25141093 & 0.75000403 & 0.59824616   \\
  24     & 1.75189751 & 1.25097723 & 0.75000277 & 0.59878935   \\
  28     & 1.75139056 & 1.25071683 & 0.75000203 & 0.59911369   \\
  32     & 1.75106290 & 1.25054825 & 0.75000154 & 0.59932298 \\ \hline
  Extrap.& 1.75002(2) &  1.25003(1) & 0.7500002(3) & 0.600003(1)   \\ \hline 
  Exact  & 1.75  & 1.25 & 0.75  & 0.6  \\ \hline
\end{tabular}
\end{table}
Of rather special nature is the state considered in Table~\ref{six}: this is
the lowest excitation from the XXZ spin-1 part of the spectrum.  As in the
antiferromagnetic regime the finite size scaling of these states can be
deducted using the spectral relation (\ref{ener-map}).  Using the same
reasoning as above the scaling dimensions $\tilde{X}_{n,m}(\gamma)$ of the
spin-1 chain in the ferromagnetically disordered regime, Eq.~(\ref{fse-fm}),
should appear doubled in the spectrum of scaling dimensions of the superspin
chain.  Comparison with (\ref{dimens}) shows that this is indeed true
(remember that $m$ in (\ref{fse-fm}) takes half-odd integer values for twist
$\varphi=\pi$):
\begin{equation}
  X_{n,n}^{m,m} = 2\tilde{X}_{n,m} = n^2\,\frac{\gamma}{2\pi} +
  m^2\,\frac{2\pi}{\gamma}\,
\end{equation}
is exactly the pure spinon part of the low energy spectrum of the mixed
superspin chain.

As an example, we have computed the finite size sequences (\ref{FSSestim}) for
the dimension $X_{0,0}^{\half,\half}(\gamma)= \frac{\pi}{\gamma}$.  This is
the state discussed at the end of Section~\ref{sec:xxz-fm}: its root
configuration changes at $\gamma=\pi/k$, $k=2,3,\ldots$ evolving into single
narrow string of $L$ roots on both levels of the Bethe ansatz for
$\gamma<\pi/L$.  

In the isotropic limit, $\gamma\to0$, this state is an $sl(2|1)$-descendent of
the (completely polarized) reference state of the superspin chain with charges
$(b,s_3)=(0,L)$ and lies outside of the low energy part of the spectrum.
Unlike the XXZ spin-1 chain the superspin chain remains conformal in the
isotropic limit of the disordered ferromagnetic regime: the holon sector,
i.e.\ states from (\ref{FSC}), (\ref{dimens}) with $m_1=-m_2$, remains
conformal for $\gamma=0$.  Taking into account (\ref{const3}) their scaling
dimensions are
\begin{equation}
\label{osp22}
  {X}_{n_1,n_2}^{m_1,-m_1}(0) = 
  \begin{cases}
    \frac{1}{4}\,(2n+1)^2+m^2 & \text{if $n_1-n_2=2n+1$ and $m_1=m$,}\\
    n^2+\frac{1}{4}\,(2m+1)^2 & \text{if $n_1-n_2=2n$ and $m_1=m+\half$}
  \end{cases}
\end{equation}
with integer $m$.  These are the conformal dimensions of the isotropic
$osp(2|2)$ spin chain \cite{JaRS03,JaSa05,GaMa07}.

We have also studied the finite size scaling behaviour of several states with
configurations which, apart from $(1,-)$ strings, contain real roots and
inifinite roots on one or both levels (the existence of such configurations in
the low energy sector of the ferromagnetic regime is indicated by our small
system analysis, see e.g.\ Table~\ref{tab:L3high}).  In all cases we have
considered these levels corresponded to descendents of the primary conformal
fields identified before, i.e.\ with scaling dimensions
$\bar{X}_{n_1,n_2}^{m_1,m_2}+n$ with integer $n$.

\section{Conclusion}
In this paper we have studied an integrable $U_q[sl(2|1)]$ vertex model built
from alternating fundamental and dual three-dimensional representations first
introduced by Gade \cite{Gade99}.  Based on its solution by means of the
algebraic Bethe ansatz we have computed the properties of this model in the
thermodynamic limit and analyzed the finite size scaling of the low energy
spectrum.  From the latter we conclude that the critical point with central
charge $c=0$ of the undeformed model identified before \cite{EsFS05} is stable
against variation of the anisotropy $\gamma$: as long as $\gamma\in[0,\pi/2)$
the ground state energy vanishes exactly without any finite size effects.  In
the continuum limit we find that the model displays a continuous spectrum of
exponents in the sector with $U_q[sl(2|1)]$-charge $b=0$.  As in the isotropic
$sl(2|1)$ superspin chain the lower edges (\ref{fse-compact}) of the continua
can be identified with the scaling dimensions in an antiperiodically twisted
spin-1 chain, in this case the Fateev-Zamolodchikov model.  The continuous
part of the conformal spectrum leads to a fine structure (\ref{fse-nonc}) in
the spectrum of the large but finite superspin chain.  This fine structure can
be explained as signature of the presence of a non-compact degree of freedom
in the continuum theory with coupling constant renormalized to an intermediate
scale of the order of the length $L$ of the superspin chain.  The
$\gamma$-dependence of the coupling constant has been identified based on our
numerical solution of the Bethe equations.  At $\gamma=\pi/2$ a level crossing
occurs changing the ground state and leading to an effective central charge
taking values $0\le c_\mathrm{eff}(\gamma)<3$ depending on the deformation
parameter $\gamma\in [\pi/2,\pi)$.  Again, this critical behaviour mirrors
that of the XXZ spin-1 chain of odd length $L$ in the antiferromagnetically
disordered regime subject to antiperiodic twisted boundary conditions.

The spectrum in the ferromagnetic regime $\pi<\gamma\le2\pi$ (or,
equivalently, that of the chain with opposite sign of the exchange constant
for anisotropies $\tilde\gamma=2\pi-\gamma$) of the superspin chain is
completely different: our finite size scaling analysis indicates that it is
the same as for the $U_q[osp(2|2)]$ spin chain with central charge $c=-1$.  It
displays separation of spin and charge degrees of freedom in the low energy
excitations with the spin part of the spectrum turning non-relativistic as
$\gamma\to2\pi$.  Unlike in the antiferromagnetic regime there are no signs of
a non-compact degree of freedom in the continuum limit: the zero charge sector
of the low energy spectrum can be identified exactly with that of the
Fateev-Zamolodchikov model in its disordered ferromagnetic phase.

In summary we have presented a comprehensive study of the critical properties
of the mixed $U_q[sl(2|1)]$ superspin chain.  The appearance of non-compact
degrees of freedom in the continuum limit of such lattice models has been
shown to be stable against deformation although it is limited to the
antiferromagnetic regime of the model.  As for the staggered six-vertex model
studied in Ref.~\onlinecite{IkJS08} our computation of the corresponding
coupling constant (\ref{fss-conj}) relies on the numerical solution of the
Bethe equations and its derivation within an analytical approach remains an
open problem.
More general phases can be expected to be found when one considers mixed
chains based on higher-dimensional representations of the superalgebra (and
its deformation).  We note, however, that already in the corresponding XXZ
spin-$S$ chains this leads to a growing number of phases (unitary and even
non-unitary) as the deformation parameter is varied
\cite{KiRe86,KiRe87,FrYF90,FrYu90}.

\acknowledgments
HF acknowledges the hospitality of the Departamento de F\'isica, UFSCar, where
much of this work has been performed.  This work has been supported by the
Deutsche Forschungsgemeinschaft, the Brazilian Foundations FAPESP and CNPq,
and the Center for Quantum Engineering and Space-Time Research (QUEST).

%
%
%
%

\end{document}